\documentclass[12pt,preprint]{aastex}

\newcommand{\ergl}{ergs~s$^{-1}$}

\newcommand{\ergcms}{ergs~cm$^{-2}$~s$^{-1}$}

\newcommand{\Chandra}{{\sl Chandra}}
\newcommand{\ROSAT}{{\sl ROSAT}}
\newcommand{\HST}{{\sl HST}}

\newcommand{\Msun}{ \ensuremath{\rm M_\odot} }
\newcommand{\ie}{{\it i.e.}}
\newcommand{\eg}{{\it e.g.}}
\newcommand{\et}{{\it et~al.}}

\slugcomment{Revision 2008.07.14}
\shorttitle{{\Chandra\ Observations of M71}}
\shortauthors{R. Elsner et al.}

\begin{document}

\title{{\Chandra\ X-ray Observatory Observations of the Globular Cluster M71}}

\author{
Ronald F. Elsner\altaffilmark{1},
Craig O. Heinke\altaffilmark{2},
Haldan N. Cohn\altaffilmark{3},
Phyllis M. Lugger\altaffilmark{3},
J. Edward Maxwell\altaffilmark{3},
Ingrid H. Stairs\altaffilmark{4},
Scott M. Ransom\altaffilmark{5},
Jason W. T. Hessels\altaffilmark{6},
Werner Becker\altaffilmark{7},
Regina H. H. Huang\altaffilmark{7},
Peter D. Edmonds\altaffilmark{8},
Jonathan E. Grindlay\altaffilmark{8},
Slavko Bogdanov\altaffilmark{8},
Kajal Ghosh\altaffilmark{9},
and
Martin C. Weisskopf\altaffilmark{9}
}

\altaffiltext{1}
{NASA Marshall Space Flight Center, VP62, Huntsville, AL 35812 (ron.elsner@nasa.gov)}
\altaffiltext{2}
{University of Virginia, Department of Astronomy, P.O. Box 400325, Charlottesville, VA 22904-4325}
\altaffiltext{3}
{Indiana University, Department of Astronomy, 727 E. Third St., Bloomington, IN 47405}
\altaffiltext{4}
{Department of Physics and Astronomy, University of British Columbia, 6224 Agricultural Road, Vancouver, BC V6T1Z1, Canada}
\altaffiltext{5}
{National Radio Astronomy Observatory, 520 Edgemont Road, Charlottesville, VA 22903}
\altaffiltext{6}
{Astronomical Institute "Anton Pannekoek", University of Amsterdam, Kruislaan 403, 1098 SJ Amsterdam, The Netherlands}
\altaffiltext{7}
{Max-Planck-Institut f\"{u}r extraterrestrische Physik, 85741 Garching bei M\"{u}nchen, Germany}
\altaffiltext{8}
{Harvard-Smithsonian Center for Astrophysics, 60 Garden St., Cambridge, MA 02138}
\altaffiltext{9}
{NASA Marshall Space Flight Center, VP62, Huntsville, AL 35812}


\begin{abstract}
We observed the nearby, low-density globular cluster M71 (NGC 6838) with the Chandra X-ray Observatory to study its faint X-ray populations.
Five X-ray sources were found inside the cluster core radius, including the known eclipsing binary millisecond pulsar (MSP) PSR J1953+1846A.
The X-ray light curve of the source coincident with this MSP shows marginal evidence for periodicity at the binary period of 4.2 h.
Its hard X-ray spectrum and luminosity resemble those of other eclipsing binary MSPs in 47 Tuc, suggesting a similar shock origin of the X-ray emission.
A further 24 X-ray sources were found within the half-mass radius, reaching to a limiting luminosity of $1.5\times10^{30}$ \ergl (0.3-8 keV).
From a radial distribution analysis, we find that $18\pm6$ of these 29 sources are associated with M71, somewhat more than predicted, and that $11\pm6$ are background sources, both galactic and extragalactic.
M71 appears to have more X-ray sources between $L_X=10^{30}$--$10^{31}$ ergs s$^{-1}$ than expected by extrapolating from other studied clusters using either mass or collision frequency.
We explore the spectra and variability of these sources, and describe the results of ground-based optical counterpart searches.
\end{abstract}


\section{Introduction\label{s:intro}}

Globular cluster X-ray sources are of interest for many reasons.
Dense globular clusters bring stars into close dynamical encounters that lead to the production of X-ray binaries (e.g. Hut, Murphy \& Verbunt 1991), and studies of globular clusters with different structural parameters can elucidate the details of these mechanisms.
Globular clusters provide concentrations of faint X-ray sources for study of X-ray populations at a known distance, age and metallicity.
This is of interest even in the least dense globular clusters, where dynamically formed X-ray binaries may be fewer than those descended from primordial binaries.
Observations of globular clusters may provide science unique to individual X-ray sources that is enabled by knowledge of the X-ray source's distance, reddening, and other properties.
An ensemble of such results for many clusters may shed light on the evolution of globular clusters and their binary populations.

Bright X-ray sources, associated with accreting neutron stars, have long been understood to be produced in globular clusters (Clark 1975).
Fainter X-ray sources, composed of combinations of accreting neutron stars in quiescence, cataclysmic variables (CVs), millisecond radio pulsars (MSPs), and chromospherically active binaries (ABs), were known in the 1980s (Hertz \& Grindlay 1983) and have been resolved with Chandra's high spatial resolution (e.g. Grindlay \et\ 2001a, Verbunt \& Lewin 2006).
A number of globular clusters have been observed with Chandra to fairly low X-ray luminosities ($<10^{32}$ \ergl).
Studies of dense clusters have identified large numbers of accreting neutron stars, CVs, ABs and MSPs (e.g. Grindlay et al. 2001a, 2001b, Pooley et al. 2002, Becker et al. 2003).
Two relatively sparse nearby clusters have been carefully studied with Chandra: M4 (Bassa et al. 2004) and NGC 288 (Kong et al. 2006).
Those studies have indicated that ABs are prevalent in sparse clusters, and that CVs are few in number, but possibly larger in number than predicted by empirical extrapolations with the density and mass of the cluster core.

M71 (NGC 6838) is of particular interest due to its close proximity to Earth (4 kpc).
A moderately short Chandra study can identify X-ray sources down to a few $10^{30}$ \ergl (0.5--2.5 keV), probing the populations of faint CVs, MSPs, and ABs.
This cluster is of moderate central density ($\rho_c = 10^{3.05} L_{\odot}$/pc$^3$) and shows no evidence for core collapse (its central concentration parameter (Djorgovski 1993) is 1.15).
It is moderately reddened (E(B-V)=0.25) and has a globular cluster metallicity ([Fe/H]=-0.73, Harris 1996, updated 2003) slightly higher than average.
The core, half-mass radius, and tidal radii are $r_c = 0.63\arcmin$, $r_h = 1.65\arcmin$, and $r_t = 8.96\arcmin$, respectively (Harris 1996, updated 2003).
Neither {\sl Einstein} (Hertz \& Grindlay 1983) nor \ROSAT\ (Verbunt 2001) detected X-ray sources obviously associated with the cluster, with the 0.5--2.5 keV \ROSAT\ HRI upper limit being $2.6 \times 10^{31}$ \ergl\ (for an assumed bremsstrahlung spectrum with temperature 0.9 keV).
The \ROSAT\ HRI did detect 10 sources in its field of view, but none of these were within the cluster half-mass radius (Verbunt 2001).

Ransom \et\ (2003, 2005) and Hessels \et\ (2007) reported the presence of a binary MSP in M71, with $P_{PSR} = 4.89$ ms and $P_{orb} = 4.24$ h, 
and the presence at 20 cm of eclipses which last roughly 20\% of the orbital period.

In this paper, we report on \Chandra\ X-ray Observatory observations of M71.
We discuss the observations, data processing, and source detection, providing source lists, in \S\ref{s:obs}, and the radial distribution of sources with respect to the nominal cluster center in \S\ref{s:dists}.
In \S\ref{s:powerlaw} we present the results of power-law spectral fits to the brighter detected X-ray sources, and in \S\ref{s:colorlum} we present and discuss the X-ray color-color diagram for these same sources.
We then discuss candidate counterparts in \S\ref{s:counterparts}, including those extracted from the 2MASS, USNO B1.0, and TYCHO-2 catalogs (\S\ref{ss:catalogs}), a recent tabulation of variable stars (\S\ref{ss:variables}), X-ray source catalogs (\S\ref{ss:xrays}), and the MSP in M71 (\S\ref{ss:msp}).
Comparison with Hubble Space Telescope (HST) observations of M71 (\S\ref{ss:hst}) will be reported in Huang \et\ (2008, in preparation).
In \S\ref{s:m71individual} we discuss spectral and temporal properties of the brightest detected sources.
A summary of our results is given in \S\ref{s:discussion}.

\section {Observations, Data Reprocessing, and Source Detection \label{s:obs}}

We obtained a 52.4-ks \Chandra\ observation (ObsID 5434) of M71 (nominal center of cluster at J2000 RA  $19^h \ 53^m \ 46.1^s$, DEC $18\arcdeg \ 46\arcmin \ 42\arcsec$) on 20--21 December 2004 using the Advanced CCD Imaging Spectrometer (ACIS) in very faint (VF), timed-exposure mode, with 3.141-s frame time.
Starting from the standard \Chandra\ X-ray Center (CXC) processing (ASCDS version number 7.6.7.1 and CALDB 3.2.1; third reprocessing) level 1 files, we reprocessed the data without applying pixel randomization.  The reprocessing included applying the current charge transfer inefficiency correction; selection of the standard ASCA grades 0, 2, 3, 4 and 6; and application of the good time filter.  We column-cleaned the data using a variant of a method developed at Pennsylvania State University\footnote{See http://www.astro.psu.edu/xray/acis/recipes/index.html}.  As is appropriate for data taken in VF mode, we used CLEAN55\footnote{See http://cxc.harvard.edu/cont-soft/software/clean55.1.0.html} to reduce the background and remove cosmic ray afterglows.
In analyzing data and unless otherwise specified, we utilized events in pulse-invariant channels corresponding to 0.3 to 8 keV.

We searched for X-ray sources in the observed field employing techniques described in Tennant (2006) which use a circular-Gaussian approximation to the point spread function (PSF).
That author gives a rather detailed account of the method in an Appendix, and there shows that for the \Chandra\ Deep Field-North (Brandt \et\ 2001) it gives results consistent with those obtained using the CIAO script wavdetect.
Using a 30 ks subset of the much longer Deep Field-North observation made it possible to know exactly which sources should be found.
The method has been used previously in other published work (\eg\ \Chandra\ observations of the globular cluster M28 reported in Becker \et\ 2003).
As a result of operating in VF mode, background levels were low throughout the observation (\eg\ $\sim 1.4 \times 10^{-7}$ counts/s/pixel over 0.3-8.0 keV for $r_{M71} \le r_h$).
Therefore within twice the M71 half-mass radius ($r_h$ = 1.65$\arcmin$), we set the signal-to-noise threshold, $S/N$, for detection to 2.0, but also required the number of source counts to be at least 5 times the statistical uncertainty in the local background estimate.
The empirical relation derived by Tennant, C$_{min} = (S/N)^2/0.81$ then implies a point-source sensitivity limit of about 4.9 counts for $r_{M71} \le 2 r_h$ and in the energy band 0.3--8.0 keV.
Because of the increase in PSF size with off-axis distance and the associated increase in background within a detection cell, for $r_{M71} > 2 r_h$ we set the S/N threshold for detection to 2.4, and again required the number of source counts to be at least 5 times the statistical uncertainty in the local background estimate.
The point-source sensitivity limit rises to about 7 counts.
Following Tennant (2006) we expect a completeness limit of about 10 counts.
We discuss the luminosities corresponding to these count sensitivity limits in \S\ref{s:dists} and \S\ref{s:powerlaw}.

The positions of X-ray sources found in this manner inside $2 r_h$ are listed in Table~\ref{t:1},
which is divided into those inside $r_h$ and those with $r_h < r_{M71} \le 2 r_h$.  The positions of both sets of sources are displayed in Figure~\ref{f:1}.
As can be seen in Figure~\ref{f:1}, the circle with radius $2 r_h$ extends slightly off the S3 CCD toward S2, which is a front-side illuminated CCD.  We find one source, {\it s55}, inside $2 r_h$ on S2 and include it in Table~\ref{t:1}.
In this table, columns 2--5 give right ascension RA (J2000), declination DEC (J2000), detect cell radius, $r_{ext}$, and the approximate number of X-ray source counts, C$_x$, detected from the source in the 0.3--8.0 keV band.
Column~6 lists the radius about the source position for inclusion of the source with 68\% confidence.
The corresponding radii for inclusion with 95\% or 99\% confidence, $r_{95}$ and $r_{99}$, are found by multiplying column~6 by 1.62 or 2.01, respectively.
Column~7 gives the distance, $r_{M71}$, of the source from the nominal center of M71.
The remaining columns have to do with candidate counterparts and are discussed in \S\ref{ss:catalogs}.

Uncertainties in the aspect solution for ACIS-S\footnote{See http://cxc.harvard.edu/cal/ASPECT/celmon/, upper left panel of the first figure} imply $\sigma_{\rm sys}\approx 0\farcs2$ (radial, $\sim0.4\arcsec$ at 90\% confidence); to be conservative, we set $\sigma_{\rm sys}=0\farcs2$ per axis.
A re-registration (boresight) analysis using the positions of the 18 2MASS candidate counterparts listed in Table~\ref{t:1} leads to the following conclusions:  (a) the existing aspect solution leads to a statistically acceptable fit with $\chi^2 \ = \ 40.4$ for 36 degrees of freedom (1 for each axis per counterpart); and (b) the best-fit changes in pointings position and roll angle are small ($\Delta\alpha = 0.03\pm0.07\arcsec$, $\Delta\delta = 0.03\pm0.07\arcsec$, and $\Delta\theta = -2.3\pm2.0\arcmin$) and, using the f-test, statistically insignificant.
Including the candidate counterpart for the MSP in M71 in this analysis does not change these conclusions.

The positions of X-ray sources detected outside $2 r_h$ are listed in Table~\ref{t:2}.  In this table, column~7 now lists the CCD on which the source was detected.  The CCD S3 is back-illuminated and therefore its response extends to lower energies than does the response of the other CCDs which are front-illuminated.
Due to dither, sources detected close to CCD boundaries may have counts on two CCDs as indicated in Table~\ref{t:2}.
The sources {\it ss33} and {\it ss34} are separated by $\sim6\arcsec$, which is significantly larger than the corresponding values for $r_{99}$ so we conclude that these are two different sources.

\section {Distributions\label{s:dists}}

M71 lies near the plane of the Galaxy, with galactic longitude $56.74\arcdeg$ and latitude $-4.56\arcdeg$, suggesting a significant contribution to the detected X-ray sources from galactic field sources as well as a contribution from extragalactic sources.
In order to determine the radial distribution of the X-ray sources detected in the S3 field, we follow the standard analysis by adopting the generalized King model profile (see, \eg, Lugger, Cohn, \& Grindlay 1995; Grindlay \et\ 2002; Heinke \et\ 2006; Lugger \et\ 2007) for the projected radial distribution function, $s(r)$, given by

\begin{equation}
s(r) = c_0 + \frac{s_0}{[ 1 + (r/r_0)^2]^\beta},
\end{equation}

\noindent where $c_0$ is the number of background field X-ray sources (galactic and extragalactic) per unit solid angle on the sky, and $s_0$ the number of globular cluster X-ray sources per unit solid angle at the cluster center.
We make the usual assumption that the projected distribution for the visible stars is given by Eq.~(1) with $\beta = 1$ and $r_0$ equal to the core radius of 0.63$\arcmin$.
We also assume the globular cluster X-ray sources are in thermal equilibrium with the stars.
Then the X-ray core radius of X-ray sources with mass $M_x = q M_0$, where $M_0$ is the nominal mass of the visible stars, is given by

\begin{equation}
r_{c,x} = (2^{1/\beta_x}-1)^{1/2} r_0,
\end{equation}

\noindent with $\beta_x = (3 q -1 ) / 2$.

In contrast to the previous analyses cited above, where the background source level was assumed to be entirely of extragalactic origin and thus known a priori, we treated the background level as a fitting parameter on an equal basis with the other parameters in Eq.~(1).
In order to determine the best-fit values for $c_{0,x}$, $s_{0,x}$, and $\beta_x$, we carried out a maximum-likelihood fit of Eq.~(1) to the radial distances of the sources from the center of M71 given in Table~\ref{t:1}, but excluded the source on S2.
We followed the procedure described in Grindlay \et\ (2002), which directly fits the unbinned radial distribution, using nonlinear optimization to maximize the likelihood.
We estimated the parameter value errors using the bootstrap method, by generating and fitting 1000 random resamplings,
with replacement, of the source radial distribution.
We took the equivalent 1-$\sigma$ error estimate for each parameter to be one half of the 68\% range about the median of the distribution of its fitted values from the bootstrap resamplings.

We note that the bootstrap method accounts for Poisson errors in the numbers of both cluster and background sources, since each resampling represents a particular realization of the underlying probability distribution defined by the original source sample.
The total number of sources included with any region of the cluster varies over the distribution of bootstrap resamplings, since an individual source may be included in each resampling 0, 1, 2, or more times.
For a distribution of 1000 resamplings of the 62-source sample within $2 r_h$, we find that the mean and standard deviation of the number of sources in the core is $5.0\pm2.2$, in close agreement with the expectation from Poisson statistics.
Thus the King-model parameter error estimates produced by the bootstrap method include the contribution from Poisson error.

We determined best-fit parameter values using all S3 sources in Table~\ref{t:1} (i.e.\ all S3 sources within $2 r_h$ of the cluster center) and also using just the subset of these sources with $C_x \ge 10$.
Our choice of $2 r_h$ for the outer radius of the fit, rather than the more common choice of $r_h$, was based on experiments with a range of values for the outer radius.
We found that the relatively small ratio of $r_h/r_c = 2.6$ for M71 required that we adopt an outer radius of at least $2 r_h$ in order to stably determine both the surface density slope, $\beta_x$, and the background level, $c_{0,x}$.
In comparison, 47~Tuc, for which Heinke \et\ (2005) adopted $r_h$ as the outer radius of the fit, has a ratio of $r_h/r_c = 7.0$.
In performing the fits, we corrected for the portion of the circular region ($r\le 2 r_h$) that lies beyond the edge of S3.
The best-fit parameters are given in Table~\ref{t:3} and the best-fit model for the radial profile is shown in Figure~\ref{f:2}.
This figure shows the fit to the observed cumulative radial profile, together with the separate cluster and background components of the model.

The large parameter uncertainties listed in Table~\ref{t:3} are the consequence of the high background source level, relative to the size of the cluster source population.
Table~\ref{t:4} compares the number of X-ray sources actually detected on CCD S3 within the radii $r_c$, $r_h$, and $2 r_h$ with the number of background (field) X-ray sources predicted by the best-fit generalized King model.
The two-sided errors on these numbers were calculated by adding in quadrature the errors propagated from the best-fit and small number Poisson errors using Eqs. (7) and (11) from Gehrels (1986).
We average the two-sided errors when they differ by $\leq 20\%$.
This results in substantial uncertainties for all parameters of the cluster source population, \eg, the size of the cluster source population within $r_h$ is $N_x = 18 \pm 6$.

Examination of Table~\ref{t:3} indicates that for the entire set of X-ray sources the best-fit value of the mass ratio is $q = 1.08\pm0.40$, suggesting that the masses of the sources are typically of order the mass of the visible stars that dominate the potential (\ie, the turnoff-mass stars).
However, we note that the large uncertainty in this parameter precludes any definitive conclusions about the typical source mass.
Similarly, the subsample of brighter X-ray sources has a larger best-fit value for the mass ratio, $q = 1.36\pm0.47$, but the two $q$ values do not differ at a statistically significant level, precluding a test of the dependence of source mass on luminosity.


We may estimate the extragalactic contribution to the background sources as follows.
Given the value for $E_{B-V}$ from Harris (1996), the relationships $A_V = 3.1 E_{B-V}$ (Rieke \& Lebofsky 1985) and
$n_H / A_V = 1.79 \times 10^{21}$ (Predehl \& Schmitt 1995) imply a value for the hydrogen column density to M71 of $n_H = 1.39 \times 10^{21}$ H atoms cm$^{-2}$.
Assuming this value for $n_H$ and a power-law spectrum with photon index 1.53 from the \Chandra\ Deep Field-South (CDF-S; Giacconi \et\ 2001), our detection limit of 4.9 counts in the 0.3--8.0 keV band corresponds to a 0.3--8.0 keV flux of $1.32 \times 10^{-15}$ \ergcms\ and a luminosity at the distance of M71 of $2.5 \times 10^{30}$ \ergl.
The corresponding value for the 0.5--2.0 keV flux is $4.2 \times 10^{-16}$ \ergcms.
Then from the CDF-S (Giacconi \et\ 2001, Eq. (1)), we find for the expectation value for the number of extragalactic sources above a 0.3--8.0 keV band detection limit of 4.9 counts of $(0.39\pm0.09) \ r_{M71}^2$ arcmin$^{-2}$ inside a radius $r_{M71}$ (in arcmin).
Repeating this exercise for a detection limit of 10 counts in the 0.3--8.0 keV band leads to an expectation value for the number of extragalactic sources of $(0.21\pm0.05)~r_{M71}^2$ arcmin$^{-2}$ inside a radius $r_{M71}$ (in arcmin).
Assuming the same value for $n_H$ but using the power-law index 1.4 from the \Chandra\ Deep Field-North (CDF-N; Brandt \et\ 2001) increases these detection limits by 3\%.


Table~\ref{t:4} also lists the number of extragalactic X-ray sources inside the radii $r_c$, $r_h$, and $2 r_h$ predicted by these results.
The errors quoted in the table were again calculated by adding in quadrature the expectation value errors and small number Poisson errors using Eqs. (7) and (11) from Gehrels (1986), resulting as before in substantial uncertainties.
Because of the large uncertainties, we cannot conclusively say whether Galactic or extra-galactic X-ray sources dominate the sources not associated with M71.
Probably both are important.
Figure~\ref{f:3} shows the distribution of the number of sources with 0.3--8.0 keV band counts greater than $C$,
$N(>C)$ vs $C$ for the sources detected inside $2 r_h$.
For $C \le 50$ the best unweighted-least-squares fit power-law index for this distribution is -0.68, while for $50 < C < 110$ the corresponding value for this index is -2.64.

\section {Power-law spectral fits\label{s:powerlaw}}


In order to estimate X-ray luminosities for the brighter sources within $2 r_h$, we carried out fits in XSPEC (Arnaud 1996) to power-law spectra for sources with at least 45 source counts in the energy band 0.3--8.0 keV plus the candidate counterpart for the MSP in M71 with 37.5 source counts, with the results shown in Table~\ref{t:5}.
For each source we constructed response files using the CIAO tool {\it mkacisrmf}, and we extracted source and local background spectral files using {\it lextrct} (Tennant 2006).
In all cases we fixed the hydrogen column density at the value, $n_H = 1.39 \times 10^{21}$ cm$^{-2}$, appropriate for M71.
In this table, column 1 gives the \Chandra\ source name from Table~\ref{t:1} (see also Figure~\ref{f:1}),
column 2 lists the minimum number of counts used to group the spectral data for fitting in XSPEC, and
columns 3--5 give the best-fit power-law index, the best-fit normalization, and the value obtained for $\chi^2$ together with the number of degrees of freedom, $\nu$.
Whenever $\chi^2/\nu \leq 2$, we provide the single parameter 67\% confidence errors calculated by XSPEC using the {\it error} command with $\Delta\chi^2 = 1$; when $\chi^2/\nu > 2$, as is the case for source {\it s52}, we do not quote any errors.
Letting the hydrogen column density vary for {\it s52} produces an improved but still unacceptable fit.
Note that the fits for the sources {\it ss03} ands {\it ss45}, while satisfying $\chi^2/\nu \leq 2$, are not acceptable with better than 99.9\% confidence.
Column~6 lists the probability of obtaining by chance that value for $\chi^2$ or greater.
Column~7 gives the corresponding unabsorbed X-ray luminosity in the bands 0.5--2.5 keV and 0.3--8.0 keV.

In order to determine X-ray flux and luminosity limits for the fainter sources within $r_h$, we divided the sources with $C_x < 45$, but not including the MSP candidate counterpart, into two groups, one with counts in the range $15 \leq C_x < 45$ (Group 1 with 7 sources), one with $C_x < 15$ (Group 2 with 15 sources).
We summed the PI spectra for each group, averaged the corresponding responses, and fit the results to power-law spectra
as above, with the results also shown in Table~\ref{t:5}.
The total number of counts for the sources in Group 2 (the faintest group) is 119.1.
Our source count sensitivity limit of 4.9 counts then corresponds to a 0.3--8.0 keV unabsorbed X-ray flux of $\sim8.0 \times 10^{-16}$ \ergcms, and a 0.3--8.0 keV unabsorbed X-ray luminosity of $\sim1.5 \times 10^{30}$ \ergl at the distance of M71.
Corresponding limits in the 0.5--2.5 keV and 0.5-6.0 keV energy bands are $\sim7.9 \times 10^{29}$ \ergl and $\sim1.1\times 10^{30}$ \ergl, respectively.
Very similar limits were found for the somewhat brighter, but fewer, sources in Group 1.

We divided the faint sources with cluster radii between $r_h$ and $2 r_h$ into two groups as above, Group 3 with $15 \leq C_x < 45$ (5 sources) and Group 4 with $C_x < 15$ (22 sources).
The summed spectra for Group 4 do not fit a power-law spectrum with $n_H = 1.39 \times 10^{21}$ cm$^{-2}$ particularly well, so we allowed $n_H$ to also vary.
We then found an acceptable fit to a power-law spectrum, with a single parameter 67\% upper limit for $n_H \leq 2 \times 10^{20}$ cm$^{-2}$.

\section {X-ray Color--Color Diagram\label{s:colorlum}}


Figure~\ref{f:4} shows an X-ray color--color diagram for sources inside $2 r_h$, using the three energy bands $S = C$(0.3-0.8 keV), $M = C$(0.8-2.0 keV), and $H = C$(2.0-8.0 keV), plotted using $(H-S)/T$ as the $x$-axis and $M/T$ as the $y$-axis, where $T = C$(0.3-8.0 keV).
We included results for {\it s52} and Group 4 on this plot, even though power-law models with $n_H$ fixed at the value appropriate for M71 did not provide acceptable fits to the X-ray spectra from these sources.
Also shown are three curves representing power-law spectra with indices ranging from -1 (right end point of curves) to -6 (left end point of curves).
The three curves are for Hydrogen column densities $n_H = 0.1$, $1.39$ (the value for M71), and $10.0$ in units of 10$^{21}$ cm$^{-2}$.

As described by Tennant (2006), using these axes all sources should lie inside the triangle defined by $S = 0$, $M = 0$, and $H = 0$.
Soft sources will lie to the left, hard sources to the right, and centrally peaked sources in the middle.
Sources to the left are more likely to be stars in our Galaxy, and sources to the right more likely to be pulsars or background AGN.
Indeed, as we describe in \S\ref{ss:catalogs}, {\it s37} and {\it s52}, which both have candidate counterparts from the 2MASS and USNO catalogs and are likely stars, lie near the line $H = 0$, while the MSP candidate counterpart {\it s08} lies near the line $S = 0$.
The candidate counterpart for {\it s52} is also listed in the Tycho-2 catalog and is undoubtedly a foreground star.

\section {Searches for counterparts\label{s:counterparts}}

\subsection {2MASS, USNO, \& TYCHO-2 catalogs\label{ss:catalogs}}

Using the {\it HEASARC BROWSE} tool in batch mode, we searched the 2MASS\footnote{See http://www.ipac.caltech.edu/2mass/}, USNO B1.0 (Monet \et\ 2003), and TYCHO-2 (Hog \et\ 2000) catalogs for possible optical and infrared counterparts.
The TYCHO-2 catalog is a subset of the USNO B1.0 catalog containing the 2.5 million brightest stars.
Thus coincidence with a TYCHO-2 catalog member potentially restricts the nature of the X-ray source.
We required that candidates lie inside the radius $r_{99}$, given in Tables~\ref{t:1} and \ref{t:2} from the Chandra X-ray source position.
The results are listed in Tables~\ref{t:1} for sources with $r_{M71} \le 2 r_h$ and Table~\ref{t:2} for sources with $2 r_h < r_{M71}$, in columns 9---11.
As noted in \S\ref{s:obs}, we carried out a re-registration (boresight) analysis using the X-ray positions and 2MASS postions for the 18 sources in Table~\ref{t:1} with 2MASS candidate counterparts.
This analysis demonstrated that for our observation there is no need to shift the on-axis pointing position or spacecraft roll angle, thus eliminating the need to search the catalogs a second time.
Column~8 lists the radial offset from the \Chandra\ position in $\arcsec$.
If there is a candidate from the 2MASS catalog, column~9 lists the quoted $J$ (the catalog also lists $H$ and $K$ magnitudes when available).
If there is a candidate from the USNO catalog, column~10 lists the $R1$ magnitude or average of $R1$ and $R2$ magnitudes (the catalog also lists $B$ and $I$ magnitudes when available).
If there is a candidate from the TYCHO-2 catalog, column~11 lists the TYCHO-2 $VT$ magnitude.
Column~12 lists the catalog and name (if there is one) for the candidate counterpart, and in a few cases other pertinent information.
Column~13 lists the probability of a chance coincidence in \%, calculated from 

\begin{equation}
P_{coinc} = N(>m) \pi r_{99}^2 / A_{search},
\end{equation}

\noindent where $N(>m)$ is the number of sources in the corresponding catalog inside the search area, $A_{search}$, with magnitude greater than that of the candidate counterpart, and $r_{99}$ the radius enclosing the X-ray source location with 99\% confidence.
For the 2MASS catalog we set $m = K$, and for the USNO B1.0 catalog $m = I$.
The number of possibilities from the TYCHO-2 catalog is very small, so $P_{coinc} << 1$ in the few cases where a potential TYCHO-2 candidate counterpart exists.
We set the search area, $A_{search}$, to the area inside the S-array boundaries for X-ray sources detected on CCDs S2, S3, or S4, and to the area inside the I-array boundaries for X-ray sources detected on CCDs I2 and I3.

As one might expect, we found no candidates from these catalogs for X-ray sources in the core of M71.
We found 18 X-ray sources with $r_c < r_{M71} \le 2 r_h$ having a single candidate counterpart in the 2MASS catalog, none with multiple counterparts.
Of these 18 X-ray sources, 3 also have a single USNO B1.0 candidate counterpart, and one ({\it s45}) has two USNO B1.0 candidate counterparts.
We found just one X-ray source, {\it s59}, having a USNO B1.0 candidate counterpart but not a candidate counterpart in the 2MASS catalog.
The X-ray source {\it s52} has a single counterpart in each of the 2MASS, USNO B1.0, and TYCHO-2 and is likely a bright foreground star.

For X-ray sources with $2 r_h < r_{M71}$, we found 24 had single candidate counterparts in the 2MASS catalog (20 of these on CCDs S2-S3-S4, 4 on I2-I3), none with multiple counterparts.
Of these 24, 16 also have a single USNO B1.0 candidate counterpart, and three have two USNO B1.0 candidate counterparts.
There are three X-ray sources with a single USNO B1.0 candidate counterpart but no 2MASS candidate counterpart, and two X-ray sources with two USNO B1.0 candidate counterparts but no 2MASS candidate counterpart.
The X-ray source {\it ss03} has a single counterpart in each of the 2MASS, USNO B1.0, and TYCHO-2 and is likely a bright foreground star.

Color-color diagrams of $J-H$ vs. $J-K$, $B-R$ vs $B$, and $R-I$ vs. $R$ show nothing unusual about the candidate counterparts to the \Chandra\ X-ray sources, except that the counterparts to {\it s52} and {\it ss03} are both bright in $B$ and $R$, as might be expected since both appear in the TYCHO-2 catalog.


\subsection {HST data\label{ss:hst}}

There are two sets of \HST\ observations of the M71 field currently in the public domain, (see HST proposal 8118, Piotto, G., also Piotto \et\ 2002; and HST proposal 10524, Ferraro, F.).
These data consist of 2183 s of WFPC2 data divided among four filters, and 628 s of ACS data.
Results from comparing these data to the \Chandra\ list of sources presented here will be reported in Huang \et\ (2008, in preparation).

\subsection {Variable sources\label{ss:variables}}

We also compared our source positions to those listed for faint variable sources in Table 9 of Park \& Nemec (2000).
Of the 23 variable sources listed, 7 lie inside a radius $2 r_h$ of the nominal center of M71 but none of them are positionally coincident with a \Chandra\ X-ray source position.
Two sources from Park \& Nemec are coincident with \Chandra\ X-ray sources outside $2 r_h$, namely {\it ss14} on CCD S4 and {\it ss52} on CCD S2.
The X-ray source {\it ss14} lies 0.88$\arcsec$ from Park \& Nemec source {\it v21}, a W UMa binary with period 0.353 d which lies in the subgiant region in the Color-Magnitude Diagram.
The X-ray source {\it ss52} lies 1.09$\arcsec$ from Park \& Nemec source {\it v19}, which may be an eclipsing binary of thus far unknown type.
They observed a rise of 0.5 magnitude over $\sim$5 h on 1996 July 12-13 (see Park \& Nemec 2000, Figure 20).
Both of these sources have 2MASS and USNO candidate counterparts (see Table~\ref{t:2}).

\subsection {X-ray catalogs\label{ss:xrays}}

As previously mentioned, Verbunt (2001) reanalyzed all the ROSAT data containing M71 and found no X-ray sources inside the cluster's half-mass radius.
Again using the {\it HEASARC BROWSE} tool in batch mode, we searched the {\it HEASARC} Master X-ray Catalog for sources within 30$\arcsec$ of a \Chandra\ X-ray source, with the results listed in Table~\ref{t:6}.
Inside $2 r_h$, one source, 1BMW 195344.3+184610, from the Brera Multi-scale Wavelength \ROSAT\ High Resolution Imager Catalog (BMWHRICAT, Panzera \et\ 2003), lies with within 1$\arcsec$ of the relatively bright \Chandra\ X-ray source {\it s05} lying just outside the M71 core radius, and within 20$\arcsec$ or less of the sources {\it s04} and {\it s06}.
We found no other coincidences inside $2 r_h$.

Outside $2 r_h$, the bright \Chandra\ X-ray {\it ss03} on CCD S4 is positionally coincident within 1$\arcsec$ of the \ROSAT\ HRI X-ray source 1RXH J195303.5+18520.
This source is probably the same as other coincidences listed in the table for {\it ss03} drawn from the ROSAT catalogs ROSPSPC, WGACAT, and BMWHRICAT.
As noted previously, this source is probably a bright foreground star.
Other coincidences within 30$\arcsec$ are listed in the table for the X-ray sources {\it ss08}, {\it ss45}, {\it is06}, and, with low probability, {\it is14}.

\subsection {The milli-second pulsar\label{ss:msp}}

The position of PSR J1953+1846A = M71A, the 4.89 ms MSP in a 4.24 h eclipsing binary in the core of M71 reported by Ransom \et\ (2003, 2005; also see Hessels \et\ 2007), is separated by 0.12$\arcsec$ from the position of the X-ray source {\it s08} in Table~\ref{t:1}, well inside the $r_{99}$ value of 0.65$\arcsec$ found for this source.
The radio timing data yield a minimum companion mass of 0.032 \Msun (Ransom \et\ 2005, Hessels \et\ 2007).
As given in Table~\ref{t:5}, its X-ray spectrum is consistent with a power-law spectrum with photon index $\gamma = 1.89\pm0.32$, and an X-ray luminosity at the distance of M71 of $6 \times 10^{30}$ \ergl\ in the 0.5--2.5 keV band and $12 \times 10^{30}$ \ergl\ in the 0.3--8.0 keV band.
The neutron star atmosphere model ({\sl nsa} in XSPEC; Zavlin \et\ 1996, Pavlov \et\ 1995) for magnetic field strengths of 0, 10$^{12}$, and 10$^{13}$ Gauss, with the distance held fixed at 4 kpc, produces bad fits, with the best null hypothesis probability being $6.2 \times 10^{-6}$.
We therefore conclude that the X-ray emission from this MSP is largely nonthermal, and is due either to magnetospheric radiation or to an intrabinary shock formed due to interaction between the relativistic pulsar wind and matter from the stellar companion (as in the eclipsing MSPs in 47 Tuc, Bogdanov \et\ 2006).

The X-ray luminosity of M71A falls inside the range of X-ray luminosities of the MSPs in the well studied globular cluster 47 Tuc (Bogdanov \et\ 2006, see their Table 4).
It also shares the non-thermal nature of its X-ray spectrum with MSPs J, O and W in 47 Tuc, with a photon power-law index of $\gamma = 1.89\pm0.32$ vs. $\sim$1---1.5 for those three MSPs.
All four of these binary MSPs show radio eclipses, and PSR J1953+1846A is also eclipsed for $\sim$20\% of its orbit (Hessels \et\ 2007).
Bogdanov, Grindlay, \& van den Berg (2005) reported variability in the X-ray emission from MSP W in 47 Tuc.
Setting zero binary phase using the radio ephemeris (Stairs \et, in preparation), and folding with 5 phase bins at the binary period of 0.1768 d = 4.2431 h = 15.2751 ks leads to the light curve shown in Figure~\ref{f:5}.
Testing this light curve using Pearson's $\chi^2$ leads to a single trial confidence level of 2.2\% for acceptance of a steady source model, providing marginal (just under 3$\sigma$) evidence in the X-ray band for periodicity at the radio binary period.
The phase spanned by the radio eclipse ($\sim$0.18---0.36, also shown in the Figure) does not quite line up with the minimum in the X-ray light curve.
The error in eclipse phase due to extrapolation from the time of the radio observations to the time of the X-ray observation is negligible.
For 47 Tuc W Bogdanov \et\ (2006) found that X-ray minimum precedes the optically derived superior conjunction of the pulsar, which may therefore also explain the displacement of the radio eclipse from the minimum in the folded X-ray light curve.
Bogdanov \et\ (2006) suggest the X-ray emission for the three eclipsing binary MSPs in 47 Tuc originates in a shock forming from the collision of a relativistic wind from the pulsar with material from its low-mass companion, a model which is as likely to apply to M71A as to the 47 Tuc sources.

\section {Individual sources\label{s:m71individual}}

Figure~\ref{f:6} displays source and local background light curves in the 0.3--8.0 keV band for the 14 X-ray \Chandra\ detected sources in Tables~\ref{t:1} and~\ref{t:2} with 0.3--8.0 keV counts $C_x > 80$, plus the source, {\it s08}, in M71's core coincident with the MSP.
Four of these sources, {\it s02}, {\it s05}, {\it s08}, and {\it s20}, lie inside the half-mass radius $r_h$.
At the timescales shown, most of these 15 sources are inconsistent with being steady at 99.9\% confidence or greater, except for {\it s08} (the source coincident with the MSP), {\it s52}, {\it ss21}, and {\it is01}.
Source {\it s02} shows a steady rise over nearly the last half of the observation. The source {\it ss06} is extremely faint for over half of our observation, suddenly flares up strongly, then decays back to a faint level, with the flare lasting lasting $\sim$10 ks.
This source has a 2MASS candidate counterpart.
The strong temporal variations in {\it s02} and {\it ss06} suggest flaring on coronally active stars.

X-ray source {\it s52} is positionally coincident with single 2MASS, USNO B1.0, and TYCHO-2 sources (see \S\ref{ss:catalogs}), and its Tycho-2 counterpart has significant proper motion.
As shown in Table~\ref{t:7}, its X-ray spectrum fits acceptably to a single MEKAL model with no interstellar absorption.
These facts suggest that {\it s52} is likely a nearby foreground star.
Source {\it s38} shows evidence for time variability (Figure~\ref{f:6}).
The USNO catalog provides morphological star/galaxy classifications, which allow us to suggest that sources {\it s45} and {\it s59} are stars while {\it s37} is a galaxy.

The brightest of the 73 X-ray sources found outside $2 r_h$, indeed the brightest in our field, is {\it ss03}.
This X-ray source is positionally coincident with single 2MASS, USNO B1.0, and TYCHO-2 sources (see \S\ref{ss:catalogs}), and on that basis is likely a nearby foreground star.
It is also coincident with what is likely a single \ROSAT\ X-ray source (see \S\ref{ss:xrays}).
The fit of its X-ray spectrum to a power-law model is statistically not very good, but a fit to a single MEKAL model is statistically far worse.
A fit to a MEKAL+MEKAL model with no interstellar absorption is an improvement over a single MEKAL model, but is still statistically worse than the fit to a power-law model (compare Tables~\ref{t:5} and \ref{t:7}).
In addition, the source is time variable, showing a definite decline during our observation.
A single MEKAL model does not fit the X-ray spectrum for source {\it ss45}, but a MEKAL+MEKAL model with no interstellar absorption does fit acceptably.
There is one 2MASS object and one USNO B1.0 object within this source's $r_{99}$ radius (see \S\ref{ss:catalogs}) as well as two \ROSAT\ HRI sources within 2--6$\arcsec$.
Table~\ref{t:7} shows that the X-ray spectrum for source {\it is06} is fit acceptably by a single MEKAL model with no interstellar absorption.
We find no counterpart in the 2MASS or USNO B1.0 catalogs, but this source is within 2.1$\arcsec$ of a faint \ROSAT\ HRI source (see \S\ref{ss:xrays}).

\section{Discussion\label{s:discussion}} 

We have found five X-ray sources (including the millisecond pulsar M71A) located within the cluster core radius.
Our radial distribution analysis indicates 1-2, and possibly all four other than M71A, are background sources.
M71A shows marginal evidence for modulation at the the binary period, slightly before the radio eclipse phase, suggesting similarity with the eclipsing millisecond radio pulsar 47 Tuc W (Bogdanov \et\ 2006).
We identify 29 X-ray sources within the half-mass radius of M71, down to a limiting X-ray luminosity (0.3-8.0 keV) $\sim 1.5 \times 10^{30}$ \ergl, of which 5-17 may be foreground or background sources.
Seven of these (all outside the core) have candidate 2MASS counterparts.
We also identify a further 108 sources outside this radius, of which the majority (71) have candidate 2MASS, USNO, or TYCHO-2 optical or near-IR counterparts.

It is of great interest to compare the populations of faint X-ray sources in different globular clusters.
Our radial distribution analysis shows that the X-ray sources associated with M71 seem to be largely confined within the half-mass radius, as seen in denser clusters (\eg\ NGC 6440, Pooley \et\ 2002).
This is in contrast to the similarly low-density cluster $\omega$ Cen, where a quiescent low-mass X-ray binary lies outside the half-mass radius (Rutledge et al. 2003).
Three likely RS CVn and eclipsing Algol stars (all cluster members) have also been astrometrically identified with Chandra or XMM-Newton sources outside the half-mass radius (Cool et al.\ 2002; Gendre, Barret, \& Webb 2003).
We suggest that this contrast may be due to the particularly low level of mass segregation in $\omega$ Cen, and its much longer half-mass relaxation time (in agreement with Verbunt \& Johnston 2000).

Studies of globular clusters show that above $L_X = 10^{31}$ \ergl, the X-ray sources seem to be dominated by CVs, while below $L_X = 10^{31}$ \ergl, ABs seem to be the dominant population (see Fig. 14 in Heinke \et\ 2005, or Fig. 6 in Kong \et\ 2006).
(Neutron stars contribute to both populations--as qLMXBs for the brighter population, and as MSPs for the fainter--but do not approach 50\% of either population in any studied cluster.)
Therefore, some constraints upon the relative frequency of ABs in different cluster environments are possible even before completion of detailed counterpart identification studies.

We count the total numbers of sources with: a) $10^{31} < L_X$; and b) $10^{30} < L_X < 10^{31}$ \ergl, in the 0.5-2.5 keV band, inside the half-mass radii of four clusters:  M4, M71, 47 Tucanae, and NGC 6397.
We calculate the numbers of background sources within each half-mass radius (except NGC 6397, where we use the 2\arcmin\ radius studied by Grindlay \et\ 2001b) predicted by the Chandra Deep Field-North studies (Brandt \et\ 2001).
For M71, we use the counts from Table 1 and derive (0.5-2.5 keV) luminosities using the spectral fit to groups 1 and 2 in Table 5, giving $L_X=1.48\times10^{29}$ ergs/s/photon for on-axis sources.  
M71's high foreground source contamination (attributable to its low Galactic latitude and low density) requires correction for foreground contamination using the results of \S\ref{s:dists}.  We find a total of $12.5\pm6.2$ X-ray sources within M71's half-mass radius between $10^{30} < L_X < 10^{31}$ ergs/s.
The background-subtracted source numbers in these $L_x$ ranges are listed in Table~\ref{t:8}, and plotted in Figure~\ref{f:7}.
The errors on the source numbers are derived from the Poisson statistics of the numbers of background sources, as in Table ~\ref{t:4}.

Figure~\ref{f:7} shows that a linear relation between scaled mass, $M_h$, inside the half-mass radius and the number, $N_h$, of X-ray sources inside the half-mass radius in the luminosity range dominated by ABs appears acceptable for most clusters.
In computing this linear fit ($N_h = a M_h$), we add in quadrature the errors on the predicted number of sources at each cluster's mass, with the actual errors on the cluster sources.
M71 stands out as having an excess of likely ABs for its mass ($12.5\pm6.2$) sources with $10^{30} < L_x < 10^{31}$ \ergl, compared to 3.5 expected), although it is less than a 2$\sigma$ excess.
Similar radial profile analyses have not yet been done for most of the other comparison clusters; if they also suffer foreground contamination, then the discrepancy will be increased.

The central density of M71 is relatively low, log $\rho_c \simeq 3.0$, suggesting that M71 may have lost fewer binaries (and hence ABs) through binary destruction mechanisms (Ivanova \et\ 2005).
Contamination of the sample by collisionally produced X-ray sources (e.g. CVs) cannot explain M71's likely abundance of X-ray sources, since M71 is a very low-density and low-Gamma cluster (Table 8).
Thorough optical identification campaigns (to reduce the uncertainties in cluster membership), and deep observations of additional globular clusters, will enable progress in understanding the formation and destruction of faint X-ray sources.



\begin{center}
 {\bf Acknowledgements}
\end{center}

This research has made use of data obtained from the High Energy Astrophysics Science Archive Center (HEASARC), provided by NASA's Goddard Space Flight Center.
COH acknowledges support from a Lindheimer Postdoctoral Fellowship at Northwestern University, and from \Chandra\ Guest Observer grants at the University of Virginia.
Those of us at NASA's Marshall Space Flight Center (MSFC) acknowledge support from the \Chandra\ Program, as well as from the \Chandra\ Guest Observer Program administered by the \Chandra\ X-ray Center.
JWTH is funded by an NSERC post-doctoral fellowship and CSA supplement.
Pulsar research at UBC is supported by an NSERC Discovery Grant.
We also thank Dr. Allyn Tennant for discussions of source finding and many aspects of \Chandra\ data analysis, as well as for sharing useful scripts.
Finally we thank the referee for several helpful comments and questions.s

\clearpage

\begin{center}
{\bf \large References}
\end{center}

\noindent {Arnaud, K.~A. 1996, in Astronomical Data Analysis Software and Systems V, ASP Conference Series, Vol. 101, Eds. Jacoby G. \& Barnes J., (San Francisco: Astronomical Society of the Pacific), p. 17} \\ \\
{Arnaud, M., \& Rothenflug, M. 1985, A\&AS, 60, 425-457} \\ \\
{Arnaud, M., \& Raymond, J. 1992, ApJ, 398, 394} \\ \\
{Bassa, C. \et\ 2004, ApJ, 609, 755} \\ \\
{Becker, W. \et\ 2003, ApJ, 594, 798} \\ \\
{Bogdanov, S., Grindlay, J.~E., \& van den Berg, M. 2005, ApJ, 630, 1029} \\ \\
{Bogdanov, S., Grindlay, J.~E., Heinke, C.~O., Camilo, F., Freire, F.~C.~C., and Becker, W. 2006, ApJ, 646, 1104} \\ \\
{Brandt, W.~N. \et\ 2001, ApJ, 122, 2810} \\ \\
{Clark, G.~W. 1975, ApJ, 199, L143} \\ \\
{Cool, A.~M., Haggard, D., Carlin, J.~L. 2002, PASP, 2002, 265, 277} \\ \\
{Djorgovski, S. 1993, in Structure and Dynamics of Globular Clusters, ASP Conference Series, Vol. 50, Eds. S.G. Djorgovski \& G. Meylan, (San Francisco: Astronomical Society of the Pacific), 373} \\ \\
{Gehrels, N. 1986, ApJ, 303, 336} \\ \\
{Gendre, B., Barret, D., \& Webb, N.~A. 2003, A\&A, 400, 521} \\ \\
{Giacconi, R. \et\ 2001, ApJ, 551, 624} \\ \\
{Grindlay, J.~E., Heinke, C., Edmonds, P.~D., \& Murray, S.~S. 2001a, Science, 292, 2290} \\ \\
{Grindlay, J.~E., Heinke, C.~O., Edmonds, P.~D., Murray, S.~S., \& Cool, A.~M. 2001b, ApJ, 563, L53} \\ \\
{Grindlay, J.~E., Camilo, F., Heinke, C.~O., Edmonds, P.~D., Gohn, H., \& Lugger, P. 2002, ApJ, 581, 470} \\ \\
{Harris, W.~E. 1996, AJ, 112, 1487} \\ \\
{Heinke, C.~O., Grindlay, J.~E., Edmonds, P.~D., Cohn, H.~N., Lugger, P.~M., Camilo, F., Bogdanov, S., \& Freire, P.~C. 2005, ApJ, 625, 796} \\ \\
{Heinke, C.~O., Wijnands, R., Cohn, H.~N., Lugger, P.~M., Grindlay, J.~E., Pooley, D., \& Lewin, W.~H.~G. 2006, ApJ, 651, 1098} \\ \\
{Hertz, P., \& Grindlay, J.~E. 1983, ApJ, 275, 105} \\ \\
{Hessels, J.~W.~T., Ransom, S.~M., Stairs, I.~H., Kaspi, V.~M., and Freire, P.~C.~C. 2007, ApJ, 670, 363} \\ \\
{Hog E. \et\ 2000, A\&A, 355, L27} \\ \\
{Hut, P, Murphy, B.~W., \& Verbunt, F. 1991, A\&A, 241, 137} \\ \\
{Ivanova, N., Rasio, F.~A., Lombardi, J.~C., Dooley, K.~L., \& Proulx, Z.~F. 2005, ApJ, 621, L109} \\ \\
{Kaastra, J.S. 1992, An X-Ray Spectral Code for Optically Thin Plasmas, Internal SRON-Leiden Report, updated version 2.0 \\ \\
{Kong, A.~K.~H., Bassa, C., Pooley, D., Lewin, W.~H.~G., Homer, L., Verbunt, F., Anderson, S.~F., \& Margon, B. 2006, ApJ, 647, 1065} \\ \\
{Liedahl, D.A., Osterheld, A.L., \& Goldstein, W.H. 1995, ApJ, 438, L115} \\ \\
{Lugger, P.~M., Cohn, H.~N., \& Grindlay, J.~E. 1995, ApJ, 439, 191} \\ \\
{Lugger, P.~M., Cohn, H.~N., Heinke, C.~O., Grindlay, J.~E., \& Edmonds, P.~E. 2007, ApJ, 657, 286} \\ \\
{Mewe, R., Gronenschild, E.H.B.M., \& van den Oord, G.H.J. 1985), A\&AS, 62, 197} \\ \\
{Mewe, R., Lemen, J.R., \& van den Oord, G.H.J. 1986, A\&AS, 65, 511} \\ \\
{Monet, D.~G. \et\ 2003, A. J., 125, 984} \\ \\
{Panzera M.R., Campana S., Covino S., Lazzati D., Mignani R.P., Moretti A., \& Tagliaferri G. 2003, A\&A, 399, 351} \\ \\
{Park, N.--K. \& Nemec, J.~M. 2000, AJ, 119, 1803} \\ \\
{Pavlov, G.G., Shibanov, Yu.A., Zavlin, V.E., \& Meyer, R.D. 1995, in “The Lives of the Neutron Stars,” ed. M.A. Alpar, U. Kiziloglu, \& J. van Paradijs (NATO ASI Ser. C, 450; Dordrecht: Kluwer), p. 71} \\ \\
{Piotto, G. \et\ 2002, A\&A, 391, 745} \\ \\
{Pooley, D. \et\ 2002, ApJ, 573, 184} \\ \\
{Pooley, D. \et\ 2003, ApJ, 591, L131} \\ \\
{Predehl, P., \& Schmitt, J.H.M.M. 1995, A\&A, 293, 889} \\ \\
{Ransom, S.~M., Hessels, J.~W.~T., Stairs, I.~H., Kaspi, V.~M., Backer, D.~C., Greenhill, L.~J., \& Lorimer, D. R. 2003 in Radio Pulsars, ASP Conference Proceedings, Vol. 302, Eds. M. Bailes, D.~J. Nice and S.~E. Thorsett, (San Francisco: Astronomical Society of the Pacific), p. 371} \\ \\
{Ransom, S., Hessels, J., Stairs, I., Kaspi, V., Freire, P., \& Backer, D. 2005, in Binary Radio Pulsars, ASP Conference Series, Vol. 328, Eds. F. A. Rasio and I. H. Stairs, (San Francisco: Astronomical Society of the Pacific), p. 199} \\ \\
{Rieke, G.~H. \& Lebofsky, M.~J. 1985, ApJ, 288, 618} \\ \\
{Rutledge, R.~E., Bildsten, L. Brown, E.~F., Pavlov, G.~G., and Zavlin, V.~E. 2003, BAAS, 35, 655} \\ \\
{Tennant, A. 2006, AJ, 132, 1372} \\ \\
{White, N.E., Giommi, P, \& Angelini, L., AAS Tucson (AZ) Jan. 1995} \\ \\
{Verbunt, F. 2001, A\&A, 368, 137} \\ \\
{Verbunt, F. \& Johnston, H.~M. 2000, A\&A, 358, 910} \\ \\
{Verbunt, F. \& Lewin, W.~H.~G. 2006, in Compact stellar X-ray sources. Eds. Walter Lewin \& Michiel van der Klis, Cambridge Astrophysics Series, No. 39, (Cambridge, UK: Cambridge University Press), pp. 341 - 379} \\ \\
{Zavlin, V.E., Pavlov, G.G., \& Shibanov, Yu.A. 1996, A\&A, 315, 141} \\

\clearpage

\begin{deluxetable}{cllrrrrrrcrrl}
\tabletypesize{\tiny}
\tablewidth{0pc}
\tablecaption{CXO X-ray sources detected with $r_{M71} \le 2 r_h$. \label{t:1}}
\tablehead{(1) & \multicolumn{1}{c}{(2)} & \multicolumn{1}{c}{(3)} & \multicolumn{1}{c}{(4)} & \multicolumn{1}{c}{(5)} & \multicolumn{1}{c}{(6)} & \multicolumn{1}{c}{(7)} & \multicolumn{1}{c}{(8)} & \multicolumn{1}{c}{(9)} & \multicolumn{1}{c}{(10)} & \multicolumn{1}{c}{(11)} & \multicolumn{1}{c}{(12)} & \multicolumn{1}{c}{(13)} } 
\startdata
 Source & \multicolumn{1}{c}{RA(J2000)} & \multicolumn{1}{c}{Dec(J2000)} & \multicolumn{1}{c}{$r_{\rm ext}$\tablenotemark{a}} & \multicolumn{1}{c}{$C_{x}$\tablenotemark{b}} & \multicolumn{1}{c}{$\sigma_x$\tablenotemark{c}} & \multicolumn{1}{c}{$r_{M71}$\tablenotemark{d}} & \multicolumn{1}{c}{Offset\tablenotemark{e}} & \multicolumn{1}{c}{$J$\tablenotemark{f}} & \multicolumn{1}{c}{$R$\tablenotemark{g}} & \multicolumn{1}{c}{$VT$\tablenotemark{h}} & \multicolumn{1}{c}{$P_{coinc}$\tablenotemark{i}} & \multicolumn{1}{l}{Comment} \\
  & \multicolumn{1}{c}{$^{\rm h}\: \ ^{\rm m}\ \ ^{\rm s}\: $\ \ \ \ } & \multicolumn{1}{c}{\ $\ \arcdeg\: \ \ \arcmin\: \ \ \arcsec$\ \ \ } & \multicolumn{1}{c}{$\arcsec$} &  & \multicolumn{1}{c}{$\arcsec$} & \multicolumn{1}{c}{$\arcmin$} & \multicolumn{1}{c}{$\arcsec$} & \multicolumn{1}{c}{mag} & \multicolumn{1}{c}{mag} &  \multicolumn{1}{c}{mag} & \multicolumn{1}{c}{\%} &  \\ \hline \hline\\[-2ex]

\multicolumn{13}{c}{Sources with $r_{M71} \le r_h$} \\ \hline

 s01 & 19 53 41.431 & 18 47 \ 9.25 & 1.5 &  4.4 & 0.52 & 1.20 &  &  &  &  &  &  \\
 s02 & 19 53 42.624 & 18 45 47.21 & 1.3 &  88.2 & 0.31 & 1.23 &  &  &  &  &  &  \\
 s03 & 19 53 43.651 & 18 47 24.69 & 1.4 &   5.6 & 0.47 & 0.92 &  &  &  &  &  &  \\
 s04 & 19 53 43.853 & 18 45 58.27 & 1.2 &  26.6 & 0.33 & 0.91 &  &  &  &  &  &  \\
 s05 & 19 53 44.389 & 18 46 10.18 & 1.2 & 301.7 & 0.30 & 0.67 &  &  &  &  &  &  \\
 s06 & 19 53 45.165 & 18 46 25.11 & 1.2 &   8.9 & 0.38 & 0.36 &  &  &  &  &  & In core \\
 s07 & 19 53 46.430 & 18 46 47.25 & 1.2 &  12.7 & 0.36 & 0.11 &  &  &  &  &  & In core \\
 s08 & 19 53 46.424 & 18 47  4.91 & 1.2 &  37.5 & 0.32 & 0.39 &  &  &  &  &  & In core, MSP \\
 s09 & 19 53 46.663 & 18 45 \ 6.05 & 1.3 & 18.4 & 0.35 & 1.61 &  &  &  &  &  &  \\
 s10 & 19 53 46.662 & 18 46 35.89 & 1.1 &  14.7 & 0.35 & 0.17 &  &  &  &  &  & In core \\
 s11 & 19 53 46.916 & 18 47 \ 9.18 & 1.2 & 20.3 & 0.34 & 0.49 &  &  &  &  &  & In core \\
 s12 & 19 53 47.211 & 18 48 \ 6.15 & 1.5 &  4.8 & 0.52 & 1.42 & 0.29 & 14.002 &  &  & 0.12  & 2MASS \\
 s13 & 19 53 47.427 & 18 46 \ 7.98 & 1.1 & 40.0 & 0.32 & 0.65 &  &  &  &  &  &  \\
 s14 & 19 53 47.453 & 18 47 59.57 & 1.5 &   6.7 & 0.46 & 1.33 &  &  &  &  &  &  \\
 s15 & 19 53 47.888 & 18 45 17.47 & 1.2 &  61.3 & 0.32 & 1.47 &  &  &  &  &  &  \\\
 s16 & 19 53 48.007 & 18 47 17.72 & 1.2 &   8.8 & 0.39 & 0.74 &  &  &  &  &  &  \\
 s17 & 19 53 48.473 & 18 47 16.39 & 1.2 &  28.6 & 0.33 & 0.80 &  &  &  &  &  &  \\
 s18 & 19 53 48.837 & 18 46 18.96 & 1.1 &  36.5 & 0.32 & 0.75 &  &  &  &  &  &  \\
 s19 & 19 53 48.849 & 18 46 34.04 & 1.1 &  56.4 & 0.32 & 0.66 & 0.29 & 15.545 &  &  & 0.58  &  2MASS \\
 s20 & 19 53 48.950 & 18 47 13.88 & 1.2 &  90.3 & 0.31 & 0.85 & 0.05 & 12.598 &  &  & 0.060 & 2MASS \\
 s21 & 19 53 49.369 & 18 45 50.58 & 1.2 &   6.1 & 0.41 & 1.16 &  &  &  &  &  &  \\
 s22 & 19 53 49.368 & 18 48 \ 0.55 & 1.5 &  6.5 & 0.47 & 1.52 & 0.48 & 15.289 &  &  & 0.88  & 2MASS \\
 s23 & 19 53 49.421 & 18 45 56.54 & 1.1 &  10.3 & 0.37 & 1.09 & 0.18 & 16.271 &  &  & 0.35  & 2MASS \\
 s24 & 19 53 49.798 & 18 46 \ 7.52 & 1.1 &  6.6 & 0.40 & 1.05 &  &  &  &  &  &  \\
 s25 & 19 53 50.702 & 18 46 55.00 & 1.2 &   6.7 & 0.41 & 1.11 &  &  &  &  &  &  \\
 s26 & 19 53 50.846 & 18 47 51.58 & 1.5 &  28.7 & 0.35 & 1.61 & 0.29 & 16.245 &  &  & 1.13  & 2MASS \\
 s27 & 19 53 51.470 & 18 46 \ 0.35 & 1.2 & 10.3 & 0.38 & 1.45 & 0.27 & 12.454 &  &  & 0.12  & 2MASS \\
 s28 & 19 53 52.719 & 18 46 35.32 & 1.3 &   6.1 & 0.44 & 1.57 &  &  &  &  &  &  \\
 s29 & 19 53 52.780 & 18 46 46.86 & 1.3 &  51.7 & 0.32 & 1.58 &  &  &  &  &  &  \\ \hline

  &  &  &  &  &  &  &  &  &  &  &  &  \\

\multicolumn{13}{c}{Sources with $r_h < r_{M71} \le 2 r_h$} \\ \hline

 s30 & 19 53 32.644 & 18 45 58.21 & 2.8 &   6.0 & 0.75 & 3.27  &  &  &  &  &  &  \\
 s31 & 19 53 32.723 & 18 46 36.74 & 2.8 &   6.9 & 0.70 & 3.17  &  &  &  &  &  &  \\
 s32 & 19 53 34.679 & 18 46 31.92 & 2.4 &   5.5 & 0.68 & 2.71  &  &  &  &  &  &  \\
 s33 & 19 53 34.876 & 18 45 \ 1.67 & 2.5 &   8.6 & 0.60 & 3.14 &  &  &  &  &  &  \\
 s34 & 19 53 37.948 & 18 44 54.57 & 2.1 &  11.1 & 0.48 & 2.64  &  &  &  &  &  &  \\
 s35 & 19 53 38.231 & 18 44 19.88 & 2.3 &   5.3 & 0.66 & 3.02  &  &  &  &  &  &  \\
 s36 & 19 53 38.252 & 18 45 40.38 & 1.8 &   7.1 & 0.51 & 2.13  &  &  &  &  &  &  \\
 s37 & 19 53 38.331 & 18 49 25.52 &  3.0 & 45.3 & 0.41 & 3.29  & 0.18 & 13.070 &  &  & 0.18 & 2MASS \\
   "  &              &             &      &      &      &      & 0.38 &  & 14.8(14.4) &  & 0.32 & USNO 1088-0475990 \\
 s38 & 19 53 38.933 & 18 45 46.61 & 1.7 & 102.7 & 0.32 & 1.94  &  &  &  &  &  &  \\
 s39 & 19 53 39.648 & 18 43 52.84 & 2.3 &  71.1 & 0.35 & 3.21  &  &  &  &  &  &  \\
 s40 & 19 53 39.943 & 18 44 25.46 & 2.0 &  64.6 & 0.34 & 2.71  &  &  &  &  &  &  \\
 s41 & 19 53 40.290 & 18 47 47.23 & 1.8 &  32.3 & 0.36 & 1.75  &  &  &  &  &  &  \\
 s42 & 19 53 41.397 & 18 44 44.55 & 1.7 &  18.9 & 0.38 & 2.26  & 0.18 & 12.676 &  &  & 0.11 & 2MASS \\
 s43 & 19 53 41.706 & 18 49 14.34 & 2.5 &   5.8 & 0.69 & 2.74  & 0.98 & 15.751 &  &  & 1.51 & 2MASS \\
 s44 & 19 53 42.119 & 18 45 11.14 & 1.5 &   4.8 & 0.51 & 1.79  &  &  &  &  &  &  \\
 s45 & 19 53 42.438 & 18 49 \ 5.27 & 2.3 &  5.8 & 0.65 & 2.54  & 0.89 & 14.866 &      &  & 1.35 & 2MASS \\
   "  &              &              &     &       &      &     & 0.51 &  & 15.1       &  &      & USNO 1088-0476353 \\
   "  &              &              &     &       &      &     & 1.10 &  & 14.8(14.5) &  & 1.14 & USNO 1088-0476355 \\
 s46 & 19 53 42.835 & 18 49 \ 1.01 & 2.2 & 14.8 & 0.46 & 2.44  &  &  &  &  &  &  \\
 s47 & 19 53 42.953 & 18 49 13.41 & 2.4 &   6.3 & 0.65 & 2.63  &  &  &  &  &  &  \\
 s48 & 19 53 43.794 & 18 44 26.65 & 1.7 &  15.2 & 0.40 & 2.33  & 0.72 & 16.071 &  &  & 1.28 & 2MASS \\
 s49 & 19 53 44.536 & 18 48 20.00 & 1.7 &   8.6 & 0.47 & 1.67  & 0.35 & 14.940 &  &  & 0.64 & 2MASS \\
 s50 & 19 53 44.967 & 18 49 52.03 & 2.8 &  31.6 & 0.43 & 3.17  &  &  &  &  &  &  \\
 s51 & 19 53 45.904 & 18 48 45.91 & 1.9 &   5.7 & 0.57 & 2.06  &  &  &  &  &  &  \\
 s52 & 19 53 47.022 & 18 44 26.78 & 1.5 &  85.9 & 0.32 & 2.27  & 0.01 & 9.812 &  &  & 0.011 &  2MASS \\
  "   &              &             &     &       &      &      & 0.08 &  & 10.5(10.5) &  & 0.035 & USNO 1087-0482780\\
   "  &              &             &     &       &      &      & 0.06 &  &  & 10.87 &  & TYC 1620-1232-1 \\
 s53 & 19 53 47.711 & 18 49 38.07 & 2.5 &  63.4 & 0.36 & 2.95  &  &  &  &  &  &  \\
 s54 & 19 53 50.931 & 18 48 28.42 & 1.8 &  27.2 & 0.37 & 2.11  & 0.65 & 13.359 &  &  & 0.16 & 2MASS \\
 s55\tablenotemark{j} & 19 53 51.292 & 18 43 47.55 & 2.0 &  23.0 & 0.39 & 3.16  &  &  &  &  &  &  \\
 s56 & 19 53 52.201 & 18 47 58.22 & 1.6 &   5.0 & 0.53 & 1.92  & 0.54 & 16.063 &  &  & 2.38 & 2MASS \\
 s57 & 19 53 52.738 & 18 47 52.18 & 1.6 &   8.8 & 0.44 & 1.95  & 0.72 & 16.018 &  &  & 1.16 & 2MASS \\
 s58 & 19 53 53.482 & 18 45 29.41 & 1.4 &   5.3 & 0.48 & 2.13  &  &  &  &  &  &  \\
 s59 & 19 53 54.710 & 18 48 38.75 & 2.2 &  12.7 & 0.47 & 2.81  & 0.81 &  & 17.8(17.5) &  & 1.85 & USNO 1088-0477382\\
 s60 & 19 53 56.921 & 18 48 16.53 & 2.2 &   8.5 & 0.55 & 3.00  &  &  &  &  &  &  \\
 s61 & 19 53 58.155 & 18 46 49.10 & 1.9 &  11.0 & 0.46 & 2.85  & 0.07 & 15.680 &  &  & 0.99 & 2MASS \\
 s62 & 19 53 58.396 & 18 47 18.34 & 2.0 &   6.1 & 0.58 & 2.97  &  &  &  &  &  &  \\
 s63 & 19 53 59.410 & 18 47 26.64 & 2.2 &   5.2 & 0.67 & 3.23  &  &  &  &  &  &  \\

\enddata

\tablenotetext{a}{Radius of the detect cell for collecting X-ray counts.}
\tablenotetext{b}{Number of 0.3--8.0 keV source counts collected in the detect cell.}
\tablenotetext{c}{Radius enclosing the true source position with 68\% confidence.  The corresponding radii for inclusion with 95\% or 99\% confidence are found by multiplying this column by 1.62 or 2.01, respectively.}
\tablenotetext{d}{Radius from the nominal center of the cluster (J2000 RA $19^h \ 53^m \ 46.1^s$, DEC +$18\arcdeg \ 46\arcmin \ 42\arcsec$).}
\tablenotetext{e}{Radial offset of candidate counterparts found by searching the HEASARC 2MASS (B/2mass; see http://www.ipac.caltech.edu/2mass/), USNO-B1.0 (I/284; Monet \et\ 2003), and TYCHO-2 (tycho2; Hog \et\ 2000) catalogs, requiring that the candidate counterpart lie within the 99\% confidence radius of a CXO source.  There are 6,414 and 3,376 2MASS sources, 15,324 and 9,196 USNO sources, and 7 and 9 TYCHO sources, appearing within the S2-S3-S4 and I2-I3 boundaries shown in Figures 3 and 4, respectively.  If blank, no candidate counterpart was found in these catalogs.}
\tablenotetext{f}{$J$ magnitude of a candidate counterpart found in the 2MASS catalog (HEASARC B/2mass; see http://www.ipac.caltech.edu/2mass/).}
\tablenotetext{g}{$R1(R2)$ magnitude of a candidate counterpart found in the USNO-B1.0 catalog (HEASARC I/284; Monet \et\ 2003).}
\tablenotetext{h}{$VT$ magnitude of a candidate counterpart found in the TYCHO-2 catalog (HEASARC tycho2; Hog \et\ 2000).}
\tablenotetext{i}{Probability of chance coincidence (see \S\ref{ss:catalogs}).}
\tablenotetext{j}{This source is on the S2 front-side CCD.  All other sources in this table are on the S3 back-side CCD.}

\end{deluxetable}

\clearpage


\begin{deluxetable}{cllrrrrrrcrrl}
\tabletypesize{\tiny}
\tablewidth{0pc}
\tablecaption{CXO X-ray sources detected with $2 r_h \le r_{M71}$. \label{t:2}}
\tablehead{(1) & \multicolumn{1}{c}{(2)} & \multicolumn{1}{c}{(3)} & \multicolumn{1}{c}{(4)} & \multicolumn{1}{c}{(5)} & \multicolumn{1}{c}{(6)} & \multicolumn{1}{c}{(7)} & \multicolumn{1}{c}{(8)} & \multicolumn{1}{c}{(9)} & \multicolumn{1}{c}{(10)} & \multicolumn{1}{c}{(11)} & \multicolumn{1}{c}{(12)} & \multicolumn{1}{c}{(13)} } 
\startdata
 Source & \multicolumn{1}{c}{RA(J2000)} & \multicolumn{1}{c}{Dec(J2000)} & \multicolumn{1}{c}{$r_{\rm ext}$\tablenotemark{a}} & \multicolumn{1}{c}{$C_{x}$\tablenotemark{b}} & \multicolumn{1}{c}{$\sigma_x$\tablenotemark{c}} & \multicolumn{1}{c}{Chip\tablenotemark{d}} & \multicolumn{1}{c}{Offset\tablenotemark{e}} & \multicolumn{1}{c}{$J$\tablenotemark{f}} & \multicolumn{1}{c}{$R$\tablenotemark{g}} & \multicolumn{1}{c}{$VT$\tablenotemark{h}} & \multicolumn{1}{c}{$P_{coinc}$\tablenotemark{i}} & \multicolumn{1}{l}{Comment} \\
  & \multicolumn{1}{c}{$^{\rm h}\: \ ^{\rm m}\ \ ^{\rm s}\: $\ \ \ \ } & \multicolumn{1}{c}{\ $\ \arcdeg\: \ \ \arcmin\: \ \ \arcsec$\ \ \ } & \multicolumn{1}{c}{$\arcsec$} &  & \multicolumn{1}{c}{$\arcsec$} & \multicolumn{1}{c}{$\arcmin$} & \multicolumn{1}{c}{$\arcsec$} & \multicolumn{1}{c}{mag} & \multicolumn{1}{c}{mag} & \multicolumn{1}{c}{mag} & \multicolumn{1}{c}{\%} &  \\ \hline \hline\\[-2ex]

 ss01 & 19 52 56.900  & 18 51 \ 8.75 & 22.7 &  49.9 & 1.97 & S4 &  &  &  &  &  &  \\
 ss02 & 19 53 \ 2.976 & 18 49 23.95 & 16.7 &  45.2 & 1.53  & S4 &  &  &  &  &  &  \\
 ss03 & 19 53 \ 3.490 & 18 51 59.52 & 19.4 & 7980.2 & 0.33 & S4 & 0.08 &  8.813 &  &  & 0.0035 & 2MASS \\
  "   &               &             &      &        &      &    & 0.14 &  & 10.7(10.6) &  & 0.0028 & USNO 1088-0473175 \\
  "   &               &             &      &        &      &    & 0.13 &  &            & 11.737 &  & TYC 1624-1644-1 \\
 ss04 & 19 53 \ 4.399 & 18 48 38.23 & 15.4 &  38.2 & 1.53  & S4 &  &  &  &  &  &  \\
 ss05 & 19 53 \ 9.151 & 18 53 50.29 & 19.1 &  85.0 & 1.29  & S4 & 1.02 & 14.360 &  &  & 3.33 & 2MASS \\
  "   &               &             &      &       &      &    &  1.21 &  & 16.2(15.8) &  & 6.33 & USNO 1088-0473600 \\
 ss06 & 19 53 10.843  & 18 50 18.08 & 13.0 &  251.6 & 0.58 & S4 & 0.20 & 13.918 &  &  & 0.50 & 2MASS \\
 ss07 & 19 53 12.205  & 18 47 59.83 & 10.6 &   32.5 & 1.16 & S4 &  &  &  &  &  &  \\
 ss08 & 19 53 14.354  & 18 51 57.83 & 13.3 &  386.1 & 0.51 & S4 &  &  &  &  &  &  \\
 ss09 & 19 53 14.636  & 18 50 23.35 & 11.2 &   25.7 & 1.36 & S4 & 1.48 & 14.577 &  &  & 3.63 & 2MASS \\
  "   &               &             &      &        &      &    & 1.92 &  & 17.3(16.6) &  & 9.20 & USNO 1088-0474040 \\
 ss10 & 19 53 16.567  & 18 46 37.74 &  8.1 &   59.9 & 0.70 & S3 & 1.08 & 15.345 &  &  & 1.82 & 2MASS \\
 ss11 & 19 53 21.991  & 18 48 39.87 &  6.6 &   35.9 & 0.73 & S3 & 1.00 & 14.916 &  &  & 1.35 & 2MASS \\
 ss12 & 19 53 24.427  & 18 49 53.59 &  6.7 &   20.6 & 0.94 & S3-S4 &  &  &  &  &  \\
 ss13 & 19 53 25.045  & 18 52 43.85 & 10.1 &   14.8 & 1.62 & S4 & 1.38 & 16.431 &  &  & 14.41 & 2MASS \\
  "   &               &             &      &        &      &    & 1.62 &  & 18.5(17.0) &  & 15.91 & USNO 1088-0474864 \\
 ss14 & 19 53 25.618  & 18 51 17.83 &  7.9 &   14.9 & 1.27 & S4 & 1.06 & 14.397 &  &  & 3.42 & 2MASS \\
  "   &               &             &      &        &      &    & 2.44 &  & 14.5(13.9) &  &  2.05 & USNO 1088-0474902 \\
 ss15 & 19 53 26.671  & 18 50 40.09 &  6.8 &   19.2 & 0.98 & S3-S4 &  &  &  &  &  \\
 ss16 & 19 53 27.949  & 18 47 \ 4.78 &  4.0 &  12.6 & 0.74 & S3 & 1.16 & 13.743 &  &  & 0.81 & 2MASS \\
  "   &               &              &      &        &   &   &    1.48 &  & 14.7(14.4) &  & 1.12 & USNO 1087-0481208 \\
 ss17 & 19 53 28.475  & 18 49 51.63 &  5.4 &   14.7 & 0.90 & S3 &  &  &  &  &  &  \\
 ss18 & 19 53 28.617  & 18 46 58.16 &  3.8 &   89.9 & 0.39 & S3 &  &  &  &  &  &  \\
 ss19 & 19 53 29.129  & 18 47 13.87 &  3.7 &   16.8 & 0.63 & S3 & 0.36 & 12.603 &  &  & 0.36 & 2MASS \\
  "   &               &             &      &        &      &     & 0.49 &   & 13.9(13.5) &  & 0.48 & USNO 1087-0481286 \\
 ss20 & 19 53 29.353  & 18 46 17.54 &  3.6 &   29.8 & 0.50 & S3 &  &  &  &  &  &  \\
 ss21 & 19 53 29.427  & 18 46 39.79 &  3.6 &   98.9 & 0.37 & S3 & 0.47 & 15.358 &  &  & 0.76 & 2MASS \\
 ss22 & 19 53 29.494  & 18 50 \ 8.47 &  5.4 &  12.6 & 0.97 & S3 &  &  &  &  &  &  \\
 ss23 & 19 53 29.996  & 18 45 19.29 &  3.5 &    7.3 & 0.84 & S3 &  &  &  &  &  &  \\
 ss24 & 19 53 30.251  & 18 51 57.63 &  7.4 &   20.1 & 1.04 & S3-S4 & 1.17 &  & 19.9(19.9) &  & 16.82 & USNO 1088-0475303 \\
 ss25 & 19 53 31.144  & 18 47 39.25 &  3.3 &   12.6 & 0.64 & S3 &  &  &  &  &  &  \\
 ss26 & 19 53 31.853  & 18 48 33.84 &  3.6 &   12.8 & 0.68 & S3 & 0.06 & 13.051 &  &  & 0.40 & 2MASS \\
  "   &               &             &      &        &    &  &    0.68 &  & 14.9(14.6) &  & 0.78 & USNO 1088-0475453 \\
 ss27 & 19 53 35.327  & 18 50 44.09 &  4.8 &    7.9 & 1.06 & S3 &  &  &  &  &  &  \\
 ss28 & 19 53 37.543  & 18 43 57.35 &  2.6 &    8.9 & 0.60 & S3 & 0.22 & 13.577 &  &  & 0.54 & 2MASS \\
 ss29 & 19 53 40.997  & 18 43 26.37 &  2.5 &   28.8 & 0.41 & S3 & 0.19 & 15.379 &  &  & 0.88 & 2MASS \\
 ss30 & 19 53 41.110  & 18 50 42.30 &  3.9 &    8.6 & 0.86 & S3 & 1.39 &  & 17.1(16.9) &  & 5.40 & USNO 1088-0476227 \\
  "   &               &             &      &        &      &    & 1.22 &  &            &  &      & USNO 1088-0476239 \\
 ss31 & 19 53 43.012  & 18 42 14.78 &  3.4 &   14.6 & 0.61 & S3 & 0.36 & 14.522 &  &  & 0.96 & 2MASS \\
  "   &               &             &      &        &      &    & 0.47 &  & 15.2(15.0) &  & 1.13 & USNO 1087-0482479 \\
 ss32 & 19 53 44.491  & 18 51 28.40 &  4.6 &   28.9 & 0.60 & S3 &  &  &  &  &  &  \\
 ss33 & 19 53 46.137  & 18 42 52.38 &  2.6 &    8.3 & 0.63 & S2-S3 & 0.53 & 14.809 &  &  & 1.09 & 2MASS \\
  "   &               &             &      &        &      &    & 0.76 &  & 15.3(15.0) &  & 1.03 & USNO 1087-0482712 \\
 ss34 & 19 53 46.572  & 18 42 55.24 &  2.6 &    8.2 & 0.62 & S2-S3 &  &  &  &  &  \\
 ss35 & 19 53 48.153  & 18 43 15.76 &  2.3 &   10.2 & 0.53 & S2-S3 &  &  &  &  &  \\
 ss36 & 19 53 48.183  & 18 50 \ 7.69 &  3.0 &  14.8 & 0.56 & S3 &  &  &  &  &  &  \\
 ss37 & 19 53 51.772  & 18 40 42.93 &  5.2 &    9.5 & 1.06 & S2 &  &  &  &  &  &  \\
 ss38 & 19 53 52.141  & 18 42 10.21 &  3.4 &   14.0 & 0.63 & S2 &  &  &  &  &  &  \\
 ss39 & 19 53 57.803  & 18 43 20.46 &  2.9 &   11.2 & 0.61 & S2 & 0.79 &  & (17.7) &  & 4.50 & USNO 1087-0483662 \\
 ss40 & 19 53 58.657  & 18 48 26.15 &  2.5 &   38.6 & 0.39 & S3 &  &  &  &  &  &  \\
 ss41 & 19 54 \ 0.169 & 18 39 29.30 &  8.1 &   44.7 & 0.79 & S2 &  &  &  &  &  &  \\
 ss42 & 19 54 \ 0.240 & 18 41 27.64 &  5.2 &   30.2 & 0.65 & S2 & 0.40 & 14.693 &  &  & 1.02 & 2MASS \\
  "   &               &             &      &        &      &    & 1.02 &  & 16.1(15.5) &  & 1.41 & USNO 1086-0482319 \\
 ss43 & 19 54 \ 0.960 & 18 43 29.97 &  3.3 &    8.2 & 0.77 & S2 &  &  &  &  &  &  \\
 ss44 & 19 54 \ 1.376 & 18 44 27.20 &  2.9 &   22.8 & 0.47 & S2 &  &  &  &  &  &  \\
 ss45 & 19 54 \ 2.852 & 18 42 50.70 &  4.3 &  797.2 & 0.32 & S2 & 0.39 & 10.654 &  &  & 0.015 & 2MASS \\
  "   &               &             &      &        &      &    & 0.62 &  & 12.3(12.2) &  & 0.053 & USNO 1087-0484047 \\
 ss46 & 19 54 \ 5.729 & 18 40 58.72 &  7.0 &   18.3 & 1.04 & S2 &  &  &  &  &  &  \\
 ss47 & 19 54 \ 6.665 & 18 42 \ 5.32 &  5.9 &   9.7 & 1.19 & S2 &  &  &  &  &  &  \\
 ss48 & 19 54 \ 6.637 & 18 42 42.25 &  5.3 &   41.6 & 0.58 & S2 &  &  &  &  &  &  \\
 ss49 & 19 54 \ 7.447 & 18 47 17.84 &  4.0 &   17.4 & 0.65 & S2 &  &  &  &  &  &  \\
 ss50 & 19 54 \ 7.856 & 18 44 13.20 &  4.5 &   10.8 & 0.88 & S2 &  &  &  &  &  &  \\
 ss51 & 19 54 \ 9.007 & 18 46 36.24 &  4.3 &   40.6 & 0.51 & S2 & 0.95 &  & 18.1(18.4) &  & 3.46 & USNO 1087-0484528 \\
 ss52 & 19 54 \ 9.562 & 18 47 22.66 &  4.7 &   16.7 & 0.75 & S2 & 0.88 & 15.866 &  &  & 2.72 & 2MASS \\
  "   &               &             &      &        &      &    & 0.70 &  & 18.0(17.7) &  & 5.56 & USNO 1087-0484568\\
 ss53 & 19 54 10.927 & 18 39 50.05 & 10.3 &    15.1 & 1.62 & S2 &  &  &  &  &  &  \\
 ss54 & 19 54 11.625 & 18 46 47.69 &  5.2 &    19.3 & 0.78 & S2 &  &  &  &  &  &  \\
 ss55 & 19 54 11.738 & 18 43 12.42 &  6.4 &    35.6 & 0.72 & S2 &  &  &  &  &  &  \\
 ss56 & 19 54 12.698 & 18 43 45.91 &  6.4 &    21.9 & 0.88 & S2 &  &  &  &  &  &  \\
 ss57 & 19 54 13.904 & 18 37 45.12 & 15.3 &    30.6 & 1.69 & S2 & 2.70 & 15.158 &  &  & 8.33 & 2MASS \\
  "   &              &             &      &         &      &    & 2.32 &        & 19.04 &  & 50.39 & USNO 1086-0483424 \\
 ss58 & 19 54 16.297 & 18 39 \ 3.75 & 13.6 &   72.7 & 1.01 & S2 & 1.15 &        & 19.9(19.7) &  &  & USNO 1086-0483626 \\
 ss59 & 19 54 18.319 & 18 43 39.81 &  8.7 &    46.6 & 0.82 & S2 &  &  &  &  &  &  \\
 is01 & 19 52 54.455 & 18 40 11.34 & 26.3 &   106.3 & 1.57 & I3 &  &  &  &  &  &  \\
 is02 & 19 53 17.086 & 18 42 34.97 &  9.7 &    25.3 & 1.20 & I3 &  &  &  &  &  &  \\
 is03 & 19 53 22.003 & 18 38 42.71 & 13.2 &    26.6 & 1.58 & I3 & 1.08 & 16.278 &  &  & 10.62 & 2MASS \\
  "   &              &             &      &         &      &    & 0.31 &  & 18.1(17.7) &  & 13.33 & USNO 1086-0479228 \\
  "   &              &             &      &         &      &    & 2.92 &  & 19.5 &  & 35.44 & USNO 1086-0479239 \\
 is04 & 19 53 24.281 & 18 33 12.71 & 27.1 &    71.4 & 1.96 & I2 & 3.60 & 16.215 &  &  & 15.03 & 2MASS \\
  "   &              &             &      &         &      &    & 3.09 &  & 16.5(15.0) &  & 4.83 & USNO 1085-0482335 \\
 is05 & 19 53 25.238 & 18 39 14.66 & 11.2 &    42.0 & 1.08 & I3 &  &  &  &  &  &  \\
 is06 & 19 53 26.354 & 18 39 55.13 &  9.6 &   144.5 & 0.57 & I3 &  &  &  &  &  &  \\
 is07 & 19 53 26.619 & 18 41 \ 3.48 &  7.8 &   26.4 & 0.97 & I3 &  &  &  &  &  &  \\
 is08 & 19 53 28.215 & 18 40 53.31 &  7.6 &    20.7 & 1.05 & I3 & 0.80 & 15.494 &  &  & 3.75 & 2MASS \\
  "   &              &             &      &         &      &    & 0.20 &  & 15.8(15.4) &  & 2.47 & USNO 1086-0479720 \\
  "   &              &             &      &         &      &    & 1.92 &  & 18.6 &  & 2.27 & USNO 1086-0479736 \\
 is09 & 19 53 28.730 & 18 40 \ 7.81 & 8.6 &     9.3 & 1.73 & I3 & 3.44 &  & 18.2 &  & 38.88 & USNO 1086-0479755 \\
  "   &              &              &      &        &      &    & 2.83 &  & 19.1(19.3) &  & 37.82 & USNO 1086-0479756 \\
 is10 & 19 53 31.432 & 18 39 40.69 &  8.6 &    12.3 & 1.52 & I2 &  &  &  &  &  &  \\
 is11 & 19 53 34.916 & 18 38 28.26 & 10.2 &    14.2 & 1.65 & I2 &  &  &  &  &  &  \\
 is12 & 19 53 38.072 & 18 37 44.80 & 11.2 &    13.1 & 1.89 & I2 &  &  &  &  &  &  \\
 is13 & 19 53 47.875 & 18 33 16.88 & 22.8 &    37.9 & 2.26 & I2 & 3.39 & 15.124 &  &  & 9.89 & 2MASS \\
  "   &              &             &      &         &      &    & 2.97 &  & (15.2) &  &  & USNO 1085-0484067 \\
  "   &              &             &      &         &      &    & 3.42 &  & 15.8(15.7) &  & 10.93 & USNO 1085-0484068 \\
 is14 & 19 53 55.822 & 18 34 51.53 & 18.3 &    65.7 & 1.40 & I2 &  &  &  &  &  &  \\

\enddata

\tablenotetext{a}{Radius of the detect cell for collecting X-ray counts.}
\tablenotetext{b}{Number of 0.3--8.0 keV source counts collected in the detect cell.}
\tablenotetext{c}{Radius enclosing the true source position with 68\% confidence.  The corresponding radii for inclusion with 95\% or 99\% confidence are found by multiplying this column by 1.62 or 2.01, respectively}
\tablenotetext{d}{CCD on which the X-ray source appears.  In a few cases, the source is dithered across the gap between CCDs.}
\tablenotetext{e}{Radial offset of candidate counterparts found by searching the HEASARC 2MASS (B/2mass; see http://www.ipac.caltech.edu/2mass/), USNO-B1.0 (I/284; Monet \et\ 2003), and TYCHO-2 (tycho2; Hog \et\ 2000) catalogs, requiring that the candidate counterpart lie within the 99\% confidence radius of a CXO source.  There are 6,414 and 3,376 2MASS sources, 15,324 and 9,196 USNO sources, and 7 and 9 TYCHO sources, appearing within the S2-S3-S4 and I2-I3 boundaries shown in Figures 3 and 4, respectively.  If blank, no candidate counterpart was found in these catalogs.}
\tablenotetext{f}{$J$ magnitude of a candidate counterpart found in the 2MASS catalog (HEASARC B/2mass; see http://www.ipac.caltech.edu/2mass/).}
\tablenotetext{g}{$R1(R2)$ magnitude of a candidate counterpart found in the USNO-B1.0 catalog (HEASARC I/284; Monet \et\ 2003).}
\tablenotetext{h}{$VT$ magnitude of a candidate counterpart found in the TYCHO-2 catalog (HEASARC tycho2; Hog \et\ 2000).}
\tablenotetext{i}{Probability of chance coincidence (see \S\ref{ss:catalogs}).}

\end{deluxetable}

\clearpage

\begin{deluxetable}{ccc}
\tabletypesize{\scriptsize}
\tablewidth{0pc}
\tablecaption{Best-fit King model parameters. \label{t:3}}
\tablehead{(1) & \multicolumn{1}{c}{(2)} & \multicolumn{1}{c}{(3)}} 
\startdata
\multicolumn{1}{c}{King model parameter} & \multicolumn{1}{c}{All 62 S3 X-ray sources} & \multicolumn{1}{c}{33 S3 sources with $C_x \ge 10$} \\ \hline\\[-2ex]

$c_{0,x}$\tablenotemark{a}  & $1.28\pm0.60$ & $0.63\pm0.31$ \\
$s_{0,x}$\tablenotemark{a}  & $6.92\pm3.42$ & $7.17\pm3.65$ \\
$\beta_x$                   & $1.13\pm0.61$ & $1.54\pm0.70$ \\
$q_x$                       & $1.08\pm0.40$ & $1.36\pm0.47$ \\
$r_{c,x} (\arcmin)$         & $0.58\pm0.26$ & $0.48\pm0.16$ \\

\enddata

\tablenotetext{a}{Units are sources per square-arcminute.}

\end{deluxetable}


\begin{deluxetable}{rcccc}
\tabletypesize{\scriptsize}
\tablewidth{0pc}
\tablecaption{Predicted No. of S3 sources inside radius $r_{M71}$ ($\arcmin$). \label{t:4}}
\tablehead{(1) & \multicolumn{1}{c}{(2)} & \multicolumn{1}{c}{(3)} & \multicolumn{1}{c}{(4)} & \multicolumn{1}{c}{(5)}} 
\startdata
\multicolumn{1}{c}{Radius\tablenotemark{a}} & \multicolumn{1}{c}{No. detected\tablenotemark{b}} & \multicolumn{1}{c}{Predicted no. of\tablenotemark{c}} & \multicolumn{2}{c}{Extragalactic contribution\tablenotemark{d}} \\
\multicolumn{1}{c}{ } & \multicolumn{1}{c}{ } & \multicolumn{1}{c}{background sources}  & \multicolumn{1}{c}{CDF-S\tablenotemark{e}}  & \multicolumn{1}{c}{CDF-N\tablenotemark{f}} \\ \hline\\[-2ex]
\multicolumn{5}{c}{All X-ray sources} \\ \hline\\[-2ex]
 $r_c = 0.63\arcmin$  & \ 5 & \ $1.6^{+2.6}_{-1.4}$ & \ $0.5^{+2.1}_{-0.5}$ & \ $0.6^{+2.1}_{-0.6}$ \\
 $r_h = 1.65\arcmin$  & 29  & $11.0\pm6.4$   & \ $3.3^{+3.1}_{-1.9}$ & \ $3.8^{+3.2}_{-2.0}$ \\
 $2 r_h = 3.30\arcmin$ & 62 & $39.4\pm19.7$\tablenotemark{g} & $10.8\pm4.6$\tablenotemark{g} & $12.2\pm4.7$\tablenotemark{g} \\ \hline\\[-2ex]
\multicolumn{5}{c}{X-ray sources with $C_x \ge 10$} \\ \hline\\[-2ex]
 $r_c = 0.63\arcmin$  & \ 4 &  \ $0.8^{+2.3}_{-0.8}$ & \ $0.3^{+2.0}_{-0.3}$ & \ $0.3^{+2.0}_{-0.3}$ \\
 $r_h = 1.65\arcmin$  & 18 &  \ $5.4^{+4.4}_{-3.5}$ & \ $1.8^{+2.6}_{-1.3}$ & \ $2.3^{+2.8}_{-1.5}$ \\
 $2 r_h = 3.30\arcmin$ & 33 & $19.4\pm10.8$\tablenotemark{g} & \ $5.8^{+3.8}_{-2.7}$\tablenotemark{g} & \ $7.5^{+4.1}_{-3.0}$\tablenotemark{g} \\

\enddata

\tablenotetext{a}{Radius in $\arcmin$.}
\tablenotetext{b}{No. of X-ray sources detected on CCD S3 inside this radius.}
\tablenotetext{c}{Predicted no. of background (field) X-ray sources on S3 inside this radius, based on the value for $c_{0,x}$ from Table~\ref{t:3}.  See the \S\ref{s:dists} text for discussion of the assigned errors.}
\tablenotetext{d}{Predicted no. of extragalactic X-ray sources on S3 inside this radius, based on results from the \Chandra\ Deep Fields North and South.  See the \S\ref{s:dists} text for discussion of the assigned errors.}
\tablenotetext{e}{Giacconi \et\ (2001).}
\tablenotetext{f}{Brandt \et\ (2001).}
\tablenotetext{g}{Reduced by the amount of area inside $2 r_h$ that falls off S3.}

\end{deluxetable}

\clearpage


\begin{deluxetable}{lccccrccc}
\tabletypesize{\scriptsize}
\tablewidth{0pc}
\tablecaption{Power-law spectral fits for selected X-ray sources. \label{t:5}}
\tablehead{(1) & \multicolumn{1}{c}{(2)} & \multicolumn{1}{c}{(3)} & \multicolumn{1}{c}{(4)} & \multicolumn{1}{c}{(5)} & \multicolumn{1}{c}{(6)} & \multicolumn{1}{c}{(7)} & \multicolumn{1}{c}{(8)} & \multicolumn{1}{c}{(8)} } 
\startdata
 \multicolumn{1}{c}{Source} & \multicolumn{1}{c}{Grouping\tablenotemark{a}} & \multicolumn{1}{c}{$n_H/10^{22}$} & \multicolumn{1}{c}{$\gamma$} & \multicolumn{1}{c}{$A$\tablenotemark{b}} & \multicolumn{1}{c}{$\chi^2/\nu$\tablenotemark{c}} & \multicolumn{1}{c}{$P(\ge \chi^2)$\tablenotemark{d}} & \multicolumn{1}{c}{$L_x$\tablenotemark{e}} & Comment \\ \hline\\[-2ex]

 \multicolumn{9}{c}{X-ray sources on S3 with $r_{M71} \le 2 r_h$\tablenotemark{f}} \\ \hline

  s02 & 15 & 0.139 & $3.09\pm0.22$ & $3.01\pm0.34$ &  5.82/ 3 & 0.12  & 15 (31)  &     \\
  s05 & 20 & 0.139 & $1.57\pm0.11$ & $7.83\pm0.62$ & 17.41/12 & 0.14  & 41 (103) &     \\
  s08 & 12 & 0.139 & $1.89\pm0.32$ & $1.12\pm0.23$ &  0.80/ 1 & 0.37  &  6 (12)  & MSP \\
  s15 & 15 & 0.139 & $2.18\pm0.24$ & $2.04\pm0.30$ &  2.10/ 2 & 0.35  & 10 (19)  &     \\
  s19 & 15 & 0.139 & $2.35\pm0.29$ & $1.95\pm0.30$ & 0.0011/ 1 & 0.97  &  9 (18)  &     \\
  s20 & 15 & 0.139 & $2.17\pm0.19$ & $3.03\pm0.37$ &  4.91/ 3 & 0.18  & 15 (29)  &     \\
  s29 & 15 & 0.139 & $1.44\pm0.26$ & $1.30\pm0.26$ &  0.020/1 & 0.89  &  7 (19)  &     \\


  s37 & 15 & 0.139 & $2.56\pm0.33$ & $1.59\pm0.26$ &  1.37/ 1 & 0.24   &  8 (14)  &    \\
  s38 & 20 & 0.139 & $1.40\pm0.18$ & $2.44\pm0.34$ &  2.60/ 2 & 0.27   & 13 (37)  &    \\
  s39 & 15 & 0.139 & $1.36\pm0.22$ & $1.75\pm0.30$ &  0.16/ 2 & 0.92   & 10 (28)  &    \\
  s40 & 15 & 0.139 & $0.20\pm0.21$ & $0.66\pm0.16$ &  1.40/ 2 & 0.50   &  6 (47)  &    \\
  s52 & 15 & 0.139 & $3.69$        & $2.21$        & 23.98/ 3 &  0.000025  & 12 (31)  & $\chi^2/\nu \geq 2$ \\
  s53 & 20 & 0.139 & $2.87\pm0.28$ & $2.43\pm0.32$ &  1.37/ 1 & 0.24   & 12 (23)  &    \\ \hline

 \multicolumn{9}{c}{Summed spectra for faint sources on S3 with $r_{M71} \le 2 r_h$} \\ \hline

 Group 1\tablenotemark{g}    & 20 & 0.139 & $1.72\pm0.14$ & $5.43\pm0.52$     &  6.37/ 7 & 0.50  & 28 (64)   &     \\
 Group 2\tablenotemark{h}    & 20 & 0.139 & $2.25\pm0.20$ & $3.99\pm0.43$     &  2.90/ 4 & 0.58  & 19 (37)  &     \\
 Groups 1+2\tablenotemark{i} & 20 & 0.139 & $1.89\pm0.11$ & $9.44\pm0.68$     &  8.34/13 & 0.82  & 47 (100)  &     \\
 Group 3\tablenotemark{j}    & 20 & 0.139 & $1.74\pm0.16$ &    &  3.30/ 4 & 0.51  &  19 (43)  &     \\
 Group 4\tablenotemark{k}    & 20 & 0.139 & $2.04$        & $5.28$            & 14.00/ 7 & 0.051 &  26 (52) & $\chi^2/\nu \geq 2$ \\
 Group 4    & 20 & $\leq 0.02$\tablenotemark{l} & $1.44\pm0.16$ & $3.32_{-0.30}^{+0.40}$ &  5.86/ 6 & 0.44 & 18 (49) &  \\ 
 Groups 3+4\tablenotemark{m} & 20 & 0.139 & $1.90\pm0.14$ & $9.07\pm0.74$     & 16.30/13 & 0.23  & 45 (96)  &     \\ \hline

 \multicolumn{9}{c}{X-ray sources with $2 r_h < r_{M71}$\tablenotemark{n}} \\ \hline

 ss03 & 20 & $0.24\pm0.013$   & $2.55\pm0.047$ & $601\pm24$ & 225.55/169 & 0.0014 & 2900 (5400) & on S4 \\
 ss06 & 20 & $0.19_{-0.076}^{+0.060}$   & $3.52_{-0.26}^{+0.49}$   & $20.54_{-5.19}^{+8.85}$   & 9.12/   9 & 0.43  & 110 (260) & on S4 \\
 ss08 & 20 & $0.29_{-0.075}^{+0.11}$    & $1.84_{-0.12}^{+0.19}$   & $20.79_{-4.22}^{+6.32}$   & 15.97/  16 & 0.46 & 100 (230) & on S4 \\
 ss45 & 20 & $0.23\pm0.031$   & $3.11\pm0.15$   & $58\pm8$   & 56.35/  31 & 0.0035 & 290 (600) & on S2 \\
 is01 & 20 & $0.17_{-0.14}^{+0.25}$      & $1.50_{-0.48}^{+0.26}$  & $5.57_{-2.16}^{+4.06}$    & 1.28/   5 & 0.94 & 30 (78) & on I3 \\
 is06 & 20 & $0.46$                      & $5.43$                  & $31.27$                   & 12.96/  4 & 0.011 & 300 (1700) & on I3; $\chi^2/\nu \geq 2$ \\

\enddata


\tablenotetext{a}{Minimum number of counts per spectral bin for fitting in XSPEC.}
\tablenotetext{b}{Power-law spectra normalization in units of $10^{-6}$.}
\tablenotetext{c}{Value of $\chi^2$ for best-fit and the number of degrees of freedom, $\nu$.}
\tablenotetext{d}{Probability of finding a value for $\chi^2$ $\ge$ the value actually found.}
\tablenotetext{e}{Unabsorbed X-ray luminosity in units of 10$^{30}$ \ergl\ assuming the distance to M71; energy bands are 0.5--2.5 (0.3--8.0) keV.}
\tablenotetext{f}{$dN/dE(E) = e^{-n_H \sigma_{ISM}} A/E^{\gamma}$ in units of photons/s-cm$^2$-keV with $E$ in keV, with $n_H$ fixed at the value $1.39 \times 10^{21}$ H atoms per cm$^2$.  The quoted errors are the one-parameter 67\% confidence level errors given by the XSPEC command {\sl error} with $\Delta\chi^2 = 1$.  Errors are not quoted when $\chi^2/\nu \geq 2$, as is the case for source s52 and Group 4 with $n_H$ fixed.}
\tablenotetext{g}{Results of fit for summed power-law spectrum for Group 1, which contains the 7 faint sources on S3 not included in the table but with $C_x \geq 15$ and with $r_{M71} \leq r_h$:  s04, s09, s11, s13, s17, s18, and s26.
From Table~\ref{t:1}, the number of source counts in this group is 199.0.}
\tablenotetext{h}{Results of power-law spectral fit for summed power-law spectrum for Group 2, which contains the 15 faintest sources with $r_{M71} \leq r_h$:  s01, s03, s06, s07, s10, s12, s14, s16, s21, s22, s23, s24, s25, s27, and s28.
From Table~\ref{t:1}, the number of source counts in this group is 119.1.}
\tablenotetext{i}{Results of fit for summed power-law spectrum for all sources in Groups 1 and 2.
The number of source counts in these groups is 318.1.} 
\tablenotetext{j}{Results of fit for summed power-law spectrum for Group 3, which contains the 5 faint sources on S3 not included in the table but with $C_x \geq 15$ and with $r_h < r_{M71} < 2 r_h$:  s41, s42, s48, s50, s54.
From Table~\ref{t:1}, the number of source counts in this group is 125.1.}
\tablenotetext{k}{Results of fit for summed power-law spectrum for Group 4, which contains the 22 faintest sources with $r_h < r_{M71} < 2 r_h$:  s30, s31, s32, s33, s34, s35, s36, s43, s44, s45, s46, s47, s49, s51, s56, s57, s58, s59, s60, s61, s62, s63.
From Table~\ref{t:1}, the number of source counts in this group is 165.1.}
\tablenotetext{l}{Since the fit for Group 4 with $n_H$ fixed is not very good, we also fit these data with $N_H$ free.  This produced an acceptable fit with a single parameter 67\% upper limit for $n_H$ of $\sim 2 \times 10^{20}$ cm$^{-2}$.}
\tablenotetext{m}{Results of fit for summed power-law spectrum for all sources in Groups 3 and 4.} 
\tablenotetext{n}{$dN/dE(E) = e^{-n_H \sigma_{ISM}} A/E^{\gamma}$ in units of photons/s-cm$^2$-keV with $E$ in keV.  The quoted errors are the one-parameter 67\% confidence level errors given by the XSPEC command {\sl error} with $\Delta\chi^2 = 1$.  Errors are not quoted when $\chi^2/\nu \geq 2$, as is the case for source is06.}

\end{deluxetable}

\begin{deluxetable}{ccllrllr}
\tabletypesize{\tiny}
\tablewidth{0pc}
\tablecaption{X-ray sources within 30$\arcsec$ of a CXO source. \label{t:6}}
\tablehead{(1) & \multicolumn{1}{c}{(2)} & \multicolumn{1}{c}{(3)} & \multicolumn{1}{c}{(4)} & \multicolumn{1}{c}{(5)} & \multicolumn{1}{c}{(6)} & \multicolumn{1}{c}{(7)} & \multicolumn{1}{c}{(8)} } 
\startdata
 Source & CCD & \multicolumn{1}{c}{Catalog\tablenotemark{a}} & \multicolumn{1}{c}{Name} & \multicolumn{1}{c}{Offset} & \multicolumn{1}{c}{RA(J2000)} & \multicolumn{1}{c}{Dec(J2000)} & \multicolumn{1}{c}{$R$\tablenotemark{b}}  \\
 &  &  &  & \multicolumn{1}{c}{$\arcsec$} &  \multicolumn{1}{c}{$^{\rm h}\: \ ^{\rm m}\ \ ^{\rm s}\: $\ \ \ \ } & \multicolumn{1}{c}{\ $\ \arcdeg\: \ \ \arcmin\: \ \ \arcsec$\ \ \ } & \multicolumn{1}{c}{10$^{-3}$ c/s} \\ \hline\\[-2ex]

 s05 & S3 & ROSAT BMWHRICAT & 1BMW 195344.3+184610 & 0.88  & 19 53 44.33  & 18 46 \ 9.98 & 0.46$\pm$0.12  \\
 s04 & S3 &                 &                      & 13.52 & 19 53 44.33  & 18 46 \ 9.98 & 0.46$\pm$0.12  \\
 s06 & S3 &                 &                      & 19.24 & 19 53 44.33  & 18 46 \ 9.98 & 0.46$\pm$0.12  \\
ss03 & S4 & ROSAT ROSHRI    & 1RXH J195303.5+18520 & 0.91  & 19 53 \ 3.5  & 18 52 \ 0.41 & 10.69$\pm$0.73 \\
     &    &                 & 1RXH J195303.2+18520 & 5.67  & 19 53 \ 3.19 & 18 52 \ 3.29 & 8.72$\pm$0.73  \\
     &    & ROSAT ROSPSPC   & 2RXP J195303.4+18515 & 3.35  & 19 53 \ 3.34 & 18 51 56.99  & 32.42$\pm$2.31 \\
     &    & ROSAT BMWHRICAT & 1BMW 195303.3+185201 & 3.69  & 19 53 \ 3.24 & 18 52 \ 0.52 & 8.89$\pm$0.56  \\
     &    &                 & 1BMW 195303.8+185201 & 4.08  & 19 53 \ 3.77 & 18 52 \ 0.59 & 10.65$\pm$0.58  \\
     &    & ROSAT WGACAT    & 1WGA J1953.0+1852    & 6.26  & 19 53 \ 3.29 & 18 52 \ 5.09 & 33.6$\pm$2.60  \\
ss08 & S4 & ROSAT BMWHRICAT & 1BMW 195314.4+185157 & 0.61  & 19 53 14.38  & 18 51 57.31  & 2.22$\pm$0.29 \\
     &    &                 & 1BMW 195314.0+185200 & 5.58  & 19 53 13.99  & 18 52 \ 0.01 & 1.46$\pm$0.23 \\
ss45 & S2 & ROSAT ROSHRI    & 1RXH J195402.9+18424 & 2.58  & 19 54 \ 2.93 & 18 42 48.35  & 1.85$\pm$0.36 \\
     &    &                 & 1RXH J195402.6+18425 & 3.71  & 19 54 \ 2.59 & 18 42 50.94  & 2.13$\pm$0.37  \\
     &    & ROSAT BMWHRICAT & 1BMW 195402.8+184249 & 2.09  & 19 54 \ 2.81 & 18 42 48.71  & 1.66$\pm$0.23  \\
     &    &                 & 1BMW 195402.5+184252 & 5.94  & 19 54 \ 2.45 & 18 42 52.2   & 2.23$\pm$0.28  \\
is06 & I3 & ROSAT BMWHRICAT & 1BMW 195326.2+183955 & 2.10  & 19 53 26.21  & 18 39 55.4   & 0.79$\pm$0.17  \\
is14 & I2 & ROSAT BMWHRICAT & 1BMW 195355.5+183516 & 24.44 & 19 53 55.49  & 18 35 15.5   & 1.74$\pm$0.35  \\

\enddata

\tablenotetext{a}{Sources in this table were extracted from the {\it HEASARC} Master X-ray Catalog, requiring that an entry lie within 30$\arcsec$ of a CXO source.  ROSHRI and ROSPSPC are the standard catalogs of pointed observations with the \ROSAT\ HRI and PSPC instruments.   WGACAT is another catalog of PSPC pointed observations analyzed using the XIMAGE tool (White \et\ 1995).  BMWHRICAT is the Brera Multi-scale Wavelet \ROSAT\ High Resolution Imager Source Catalog (BMW-HRI) (Panzera \et\ 2003), derived from all \ROSAT\ HRI pointed observations with exposure time longer than 100 seconds available in the \ROSAT\ public archives, and analyzed using a wavelet detection algorithm.}
\tablenotetext{b}{Measured count rate.}

\end{deluxetable}



\begin{deluxetable}{cccccccrcc}
\tabletypesize{\tiny}
\tablewidth{0pc}
\tablecaption{MEKAL and MEKAL+MEKAL spectral fits. \label{t:7}}
\tablehead{(1) & \multicolumn{1}{c}{(2)} & \multicolumn{1}{c}{(3)} & \multicolumn{1}{c}{(4)} & \multicolumn{1}{c}{(5)} & \multicolumn{1}{c}{(6)} & \multicolumn{1}{c}{(7)} & \multicolumn{1}{c}{(8)} & \multicolumn{1}{c}{(9)} & \multicolumn{1}{c}{(10)} } 
\startdata
 Source  & Model\tablenotemark{a}  & \multicolumn{1}{c}{Grouping\tablenotemark{b}} & \multicolumn{1}{c}{$T_1$} & \multicolumn{1}{c}{$A_1 / 10^{6}$} & \multicolumn{1}{c}{$T_2$} & \multicolumn{1}{c}{$A_2 / 10^{6}$} & \multicolumn{1}{c}{$\chi^2/\nu$\tablenotemark{c}} & \multicolumn{1}{c}{$P(\ge \chi^2)$\tablenotemark{d}} & \multicolumn{1}{c}{$F_x$\tablenotemark{e}} \\ \hline\\[-2ex]

  s52 & MEKAL       & 15 & $0.41\pm0.06$ & $2.17\pm0.29$ &               &           & 0.81/ 3 & 0.85     &   10(11)   \\
 ss03 & MEKAL+MEKAL & 20 & $4.00\pm0.21$ & $790_{20}^{30}$ & $0.96\pm0.06$ & $49\pm12$ & 322/165 & 3.37e-12 & 1470(2800) \\
 ss45 & MEKAL+MEKAL & 20 & $0.83\pm0.06$ & $9.9\pm1.8$   & $2.84\pm0.33$ & $51\pm4$  & 28.1/30 & 0.56     &  130(200)  \\
 is06 & MEKAL       & 20 & $1.03\pm0.08$ & $8.12\pm0.90$ &               &           & 6.12/ 5 & 0.29     &    29(34)  \\

\enddata

\tablenotetext{a}{The XSPEC MEKAL model calculates the emission from a hot diffuse gas at temperature T and with normalization A as given at http://heasarc.gsfc.nasa.gov/docs/xanadu/xspec/manual/XSmodelMekal.html (Mewe, Gronenschild, \& van den Oord 1985; Mewe, Lemen, \& van den Oord 1986; Kaastra 1992; Liedahl, Osterheld, \& Goldstein 1995; Arnaud \& Rothenflug 1985; Arnaud \& Raymond 1992).
The quoted errors are the one-parameter 67\% confidence level errors given by the XSPEC command {\sl error} with $\Delta\chi^2 = 1$.}
\tablenotetext{b}{Minimum number of counts per spectral bin for fitting in XSPEC.}
\tablenotetext{c}{Value of $\chi^2$ for best-fit and the number of degrees of freedom, $\nu$.}
\tablenotetext{d}{Probability of finding a value for $\chi^2$ $\ge$ the value actually found.}
\tablenotetext{e}{X-ray luminosity in units of 10$^{30}$ \ergl; energy bands are 0.5--2.5 (0.3--8.0) keV.}

\end{deluxetable}



\begin{deluxetable}{ccccc}
\tabletypesize{\scriptsize}
\tablewidth{0pc}
\tablecaption{Background-subtracted source numbers for selected globular clusters. \label{t:8}}
\tablehead{\multicolumn{1}{c}{(1)} & \multicolumn{1}{c}{(2)} & \multicolumn{1}{c}{(3)} & \multicolumn{1}{c}{(4)} & \multicolumn{1}{c}{(5)} } 
\startdata
   & \multicolumn{1}{c}{M4} & \multicolumn{1}{c}{M71} & \multicolumn{1}{c}{47 Tuc} & \multicolumn{1}{c}{NGC 6397} \\ \hline\\[-2ex]

Distance (kpc)                  & 1.73 & 4.0  &  4.5 & 2.55  \\
$\Gamma$\tablenotemark{a}       & 1    & 0.11 & 24.9 & 2.52  \\
$M{_h}$\tablenotemark{b}        & 1    & 0.30 & 10.2 & 0.78 \\
$10^{31} < L_x\mbox{(\ergl)\tablenotemark{c}}$ & 1\tablenotemark{d} & 2.2 & 23 & 8 \\
$10^{30} < L_x\mbox{(\ergl)\tablenotemark{c}} < 10^{31}$ & 6.6 & 12.5 & 116 & 7 \\
Error & 3.0 & 6.2 & 3.9 & 2.4 \\
References\tablenotemark{e} & 1 & 2 & 3 & 4 \\


\enddata

\tablenotetext{a}{Collision number, $\Gamma \propto \rho_0^{1.5} r_c^2$, normalized to the value for M4, where $\rho_0$ is the cluster central density and $r_c$ the core radius.
Cluster parameters come from Harris (1996; updated Feb. 2003), except M4 for which $\rho_0$ and $M_V$ are computed from the distance and reddening of Richer \et\ (1997).}
\tablenotetext{b}{Scaled cluster mass $M_h$ inside $r_h$.
Following Kong \et\ (2006), scaled values for $M_h$ were calculated from $M_h = 10^{-0.4 (M_V - M_{V,M4})}$, where $M_V$ is the cluster absolute visual magnitude from Harris (1996, updated 2003).
The value $M_{V,M4}$ for the cluster M4 is -7.2.
}
\tablenotetext{c}{$L_x$ is the X-ray luminosity of individual sources in the 0.5---2.5 keV energy band}
\tablenotetext{d}{M4 possesses one optically identified X-ray source and likely CV with $L_x > 10^{31}$ \ergl.}
\tablenotetext{e}{For each cluster, the basic data for this table were extracted from: 1 Bassa \et\ 2004; 2 this work; 3 Heinke \et\ 2005, but correcting for a distance of 4.5 kpc; 4 Grindlay \et\ 2001b.}

\end{deluxetable}

\begin{figure}[htbp]
\epsscale{1.0}
\includegraphics[scale=0.90,angle=270]{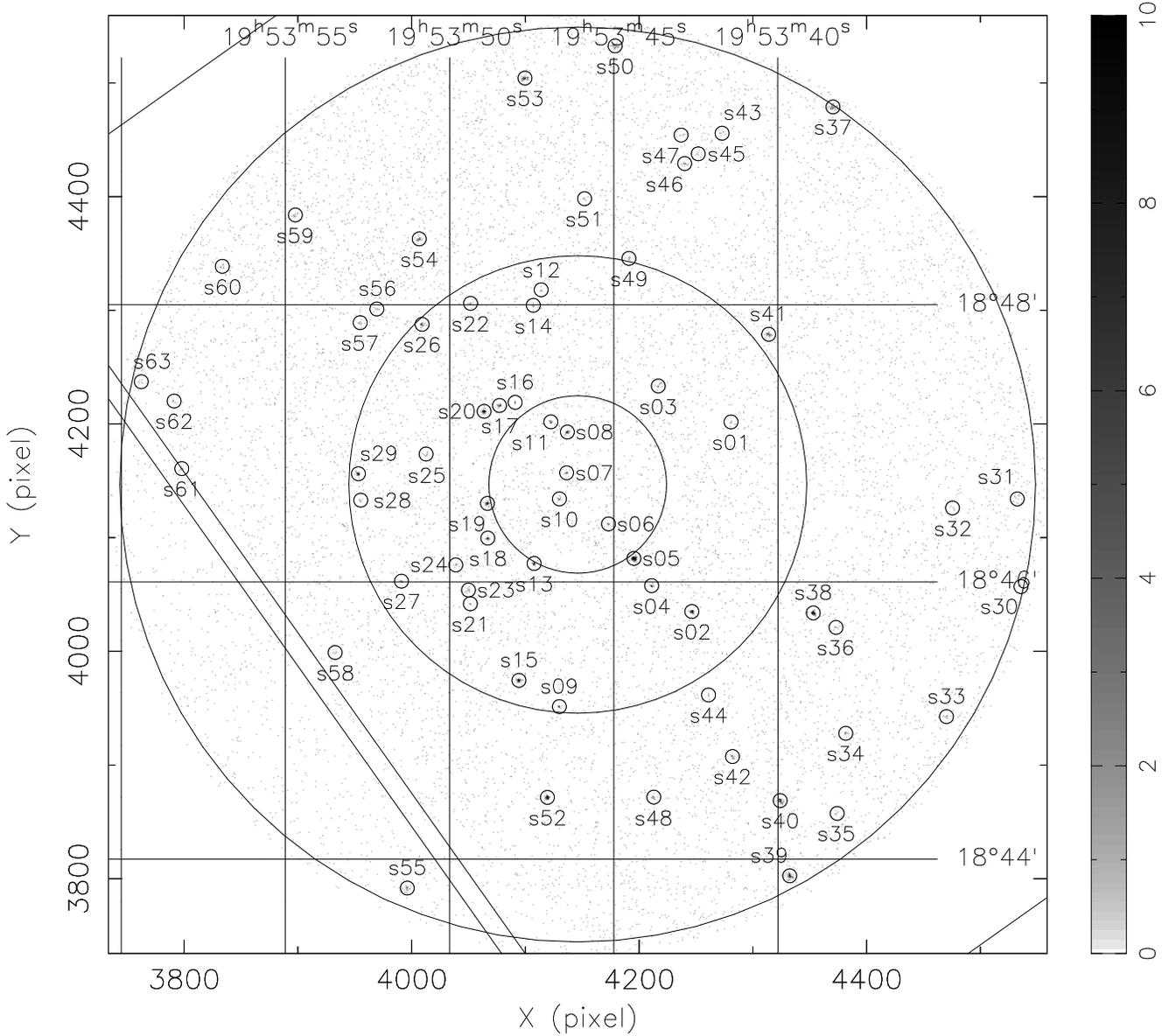}
\caption{\label{f:1}
ACIS-S S3 0.3--8.0 keV image of the globular cluster M71 for $r_{M71} \le 2 r_{h}$.  The small circles show the positions of the 29 X-ray sources listed in Table~\ref{t:1} with $r_{M71} \le r_{h}$ and the 34 X-ray sources also listed in Table~\ref{t:1} with $r_{h} < r_{M71} \le 2 r_{h}$.  The large circles are centered on the nominal center of the cluster and have radii $r_c$ (inner circle), $r_h$ (middle circle), and $2r_h$ (outer circles).  Straight lines mark the nominal boundaries of the S3 and S2 CCDs, with most of the figure falling on S3 and the lower left hand portion on S2.
}
\label{f:fig1}
\end{figure}



\begin{figure}[htbp]
\epsscale{1.0}
\plotone{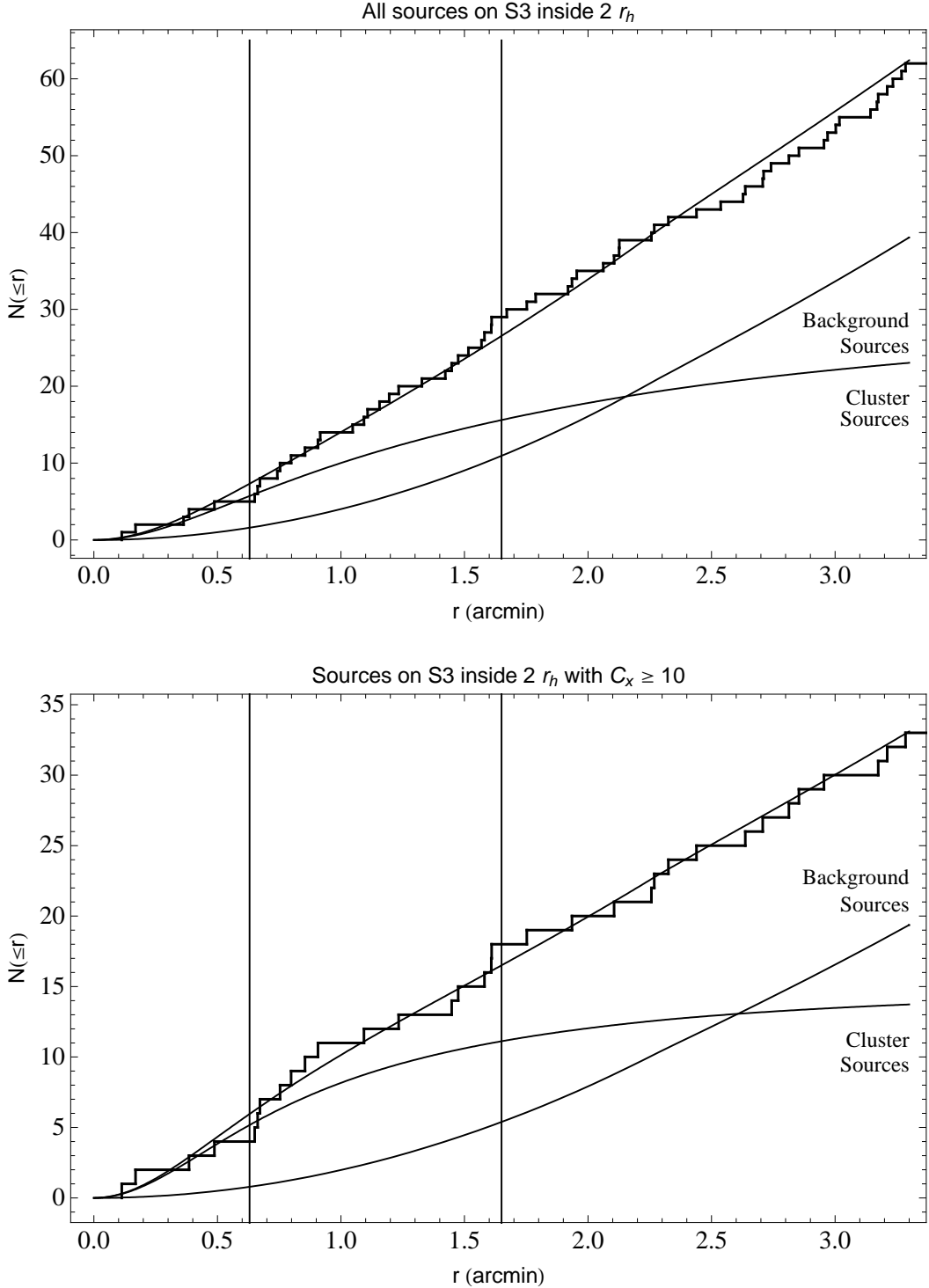}
\caption{\label{f:2}
Number of sources, $N(\leq r_{M71})$, inside radius $r_{M71} \ (\arcmin)$ vs. $r_{M71}$.
The top panel is for all S3 sources inside $2 r_h$, and the bottom panel for those sources inside $2 r_h$ with numbers of 0.3--8.0 keV counts $C_x \ge 10$.
The histogram shows the actual number of detected sources inside the corresponding radius, while the upper solid curve shows the number predicted by the best-fit King model.
The lower curves are the model predicted background and cluster contributions, as indicated.
The vertical lines mark the core radius, $r_c$, and half-mass radius, $r_h$.
}
\label{f:fig2}
\end{figure}



\begin{figure}[htbp]
\epsscale{1.0}
\plotone{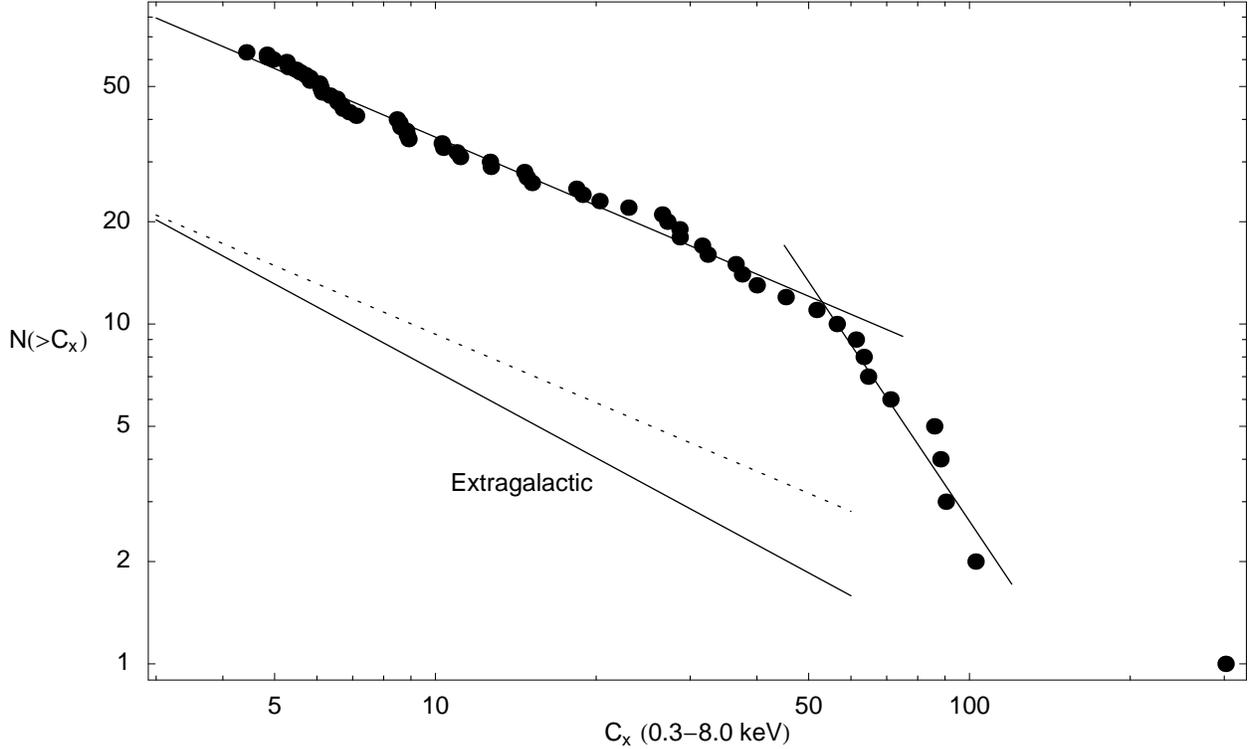}
\caption{\label{f:3}
The M71 $\log{N}$--$\log{C_x}$ distribution for sources with $r_{M71} \le 2 r_h$.  The solid lines through the data represent unweighted least-squares fits to a power-law for $C_x$(0.3--8.0 keV) $< 50$ and $C_x$(0.3--8.0 keV) $> 50$.  The functions are $N(>C_x) = 10^a / \ C_x^b$, with $(a, b) = $ (2.22, 0.67) for $C_x$(0.3--8.0 keV) $< 50$, and $(a, b) = $ (5.10, 2.34) for $C_x$(0.3--8.0 keV) $> 50$.
The lower lines show the estimated contribution from extragalactic sources (solid Giacconi \et\ 2001, dotted Brandt \et\ 2001).
}
\label{f:fig3}
\end{figure}



\begin{figure}[htbp]
\epsscale{1.0}
\plotone{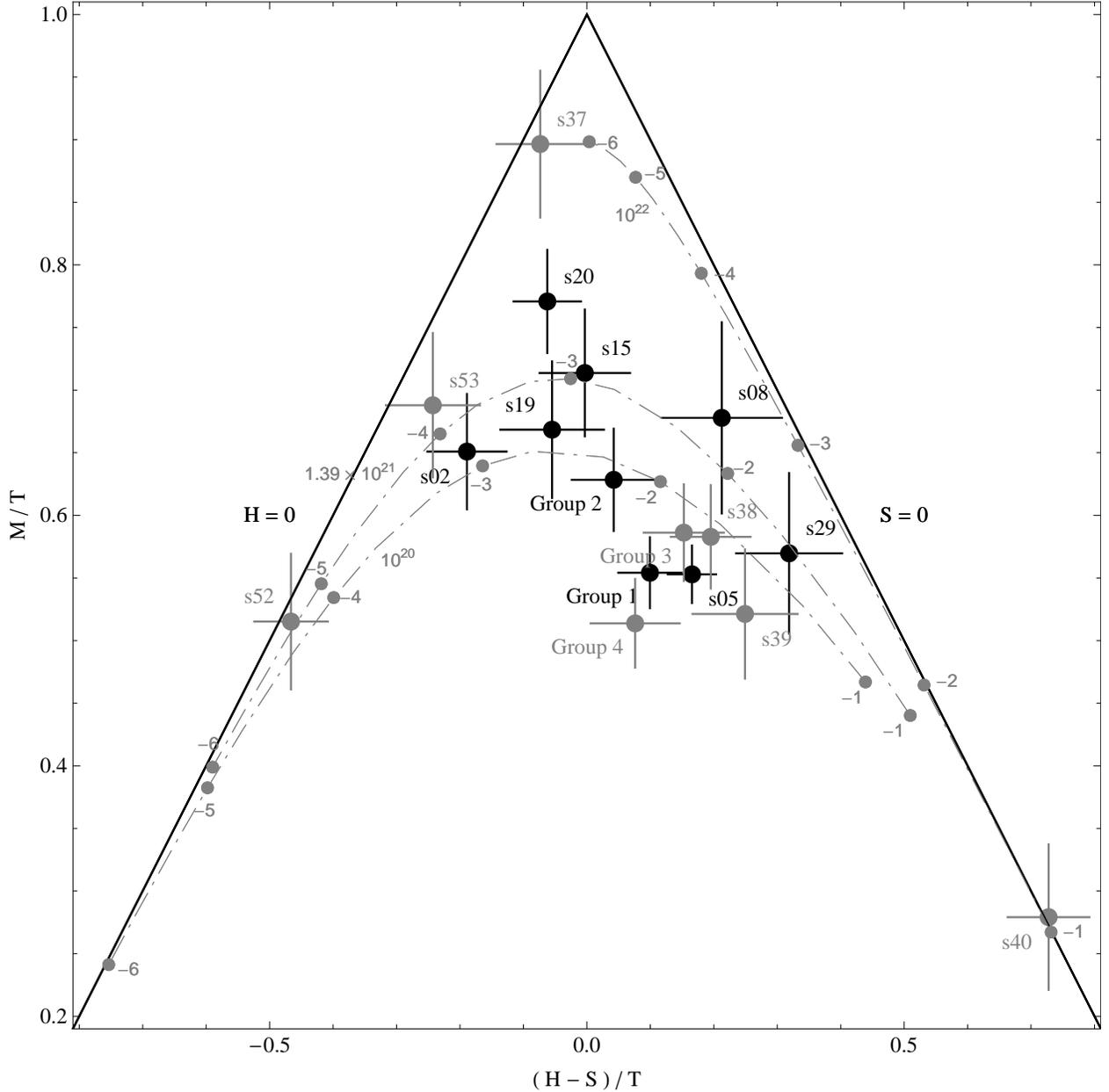}
\caption{\label{f:4}
Color-color diagram for sources with $r_{M71} \le 2 r_{h}$ having 45 or more source counts in the energy band $T =$ 0.3--8.0 keV, plus {\it s08}, the MSP candidate counterpart, and the summed spectra for Groups 1--4.
Sources inside $r_h$ are shown with large black dots while sources between $r_h$ and $2 r_h$ are shown with large grey dots.
The bands are $S =$ 0.3--0.8 keV, $M =$ 0.8--2.0 keV, and $H =$ 2.0--8.0 keV.
The dot-dashed curves are the PIMMS predicted values for power-law spectra with indices ranging from -1 (hardest) to -6 (softest), and column densities $n_H = 1.0 \times 10^{20}$ (bottom), $1.39 \times 10^{21}$ (middle), and $1.0 \times 10^{22}$ (top) cm$^{-2}$.
The labeled small grey dots show the positions along these curves of the photon indices -1 to -6, from right (hardest) to left (softest), in increments of -1.
Although we have included results for {\it 52} and Group 4 on this plot, power-law models with $n_H$ fixed at the value appropriate for M71 did not provide acceptable fits to the X-ray spectra from these sources.
}
\label{f:fig4}
\end{figure}



\begin{figure}[htbp]
\epsscale{1.0}
\plotone{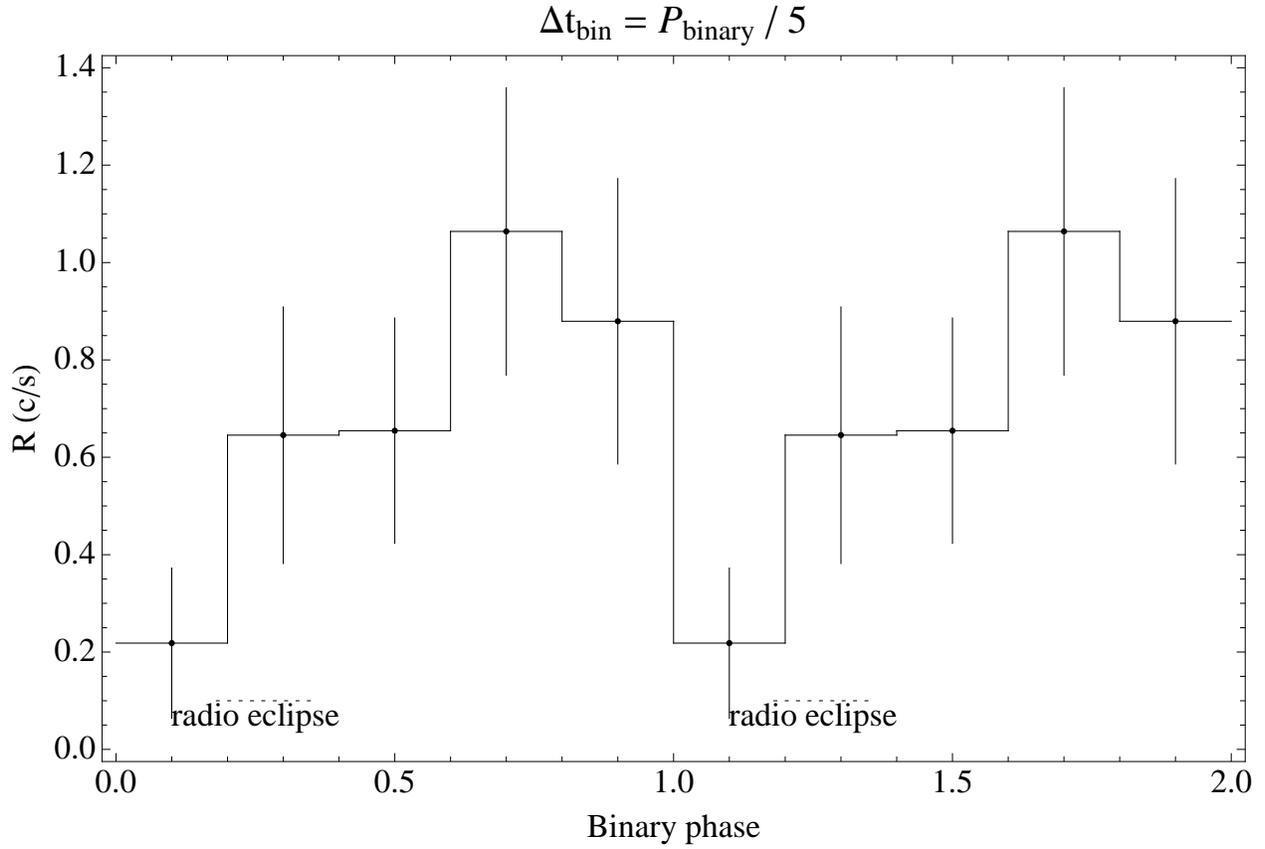}
\caption{\label{f:5}
The 0.3---8.0 keV light curve for PSR J1953+1846A, folded at the reported radio binary period of 0.1768 d = 4.2431 h = 15.2751 ks.
The bin size is 1/5 the binary period.
The single period probability of chance occurrence of the corresponding value of $\chi^2$ is 2.2\%.
Also shown is the phase spanned by the radio eclipse.}
\label{f:fig5}
\end{figure}

\begin{figure}[htbp]
\epsscale{1.0}
\plotone{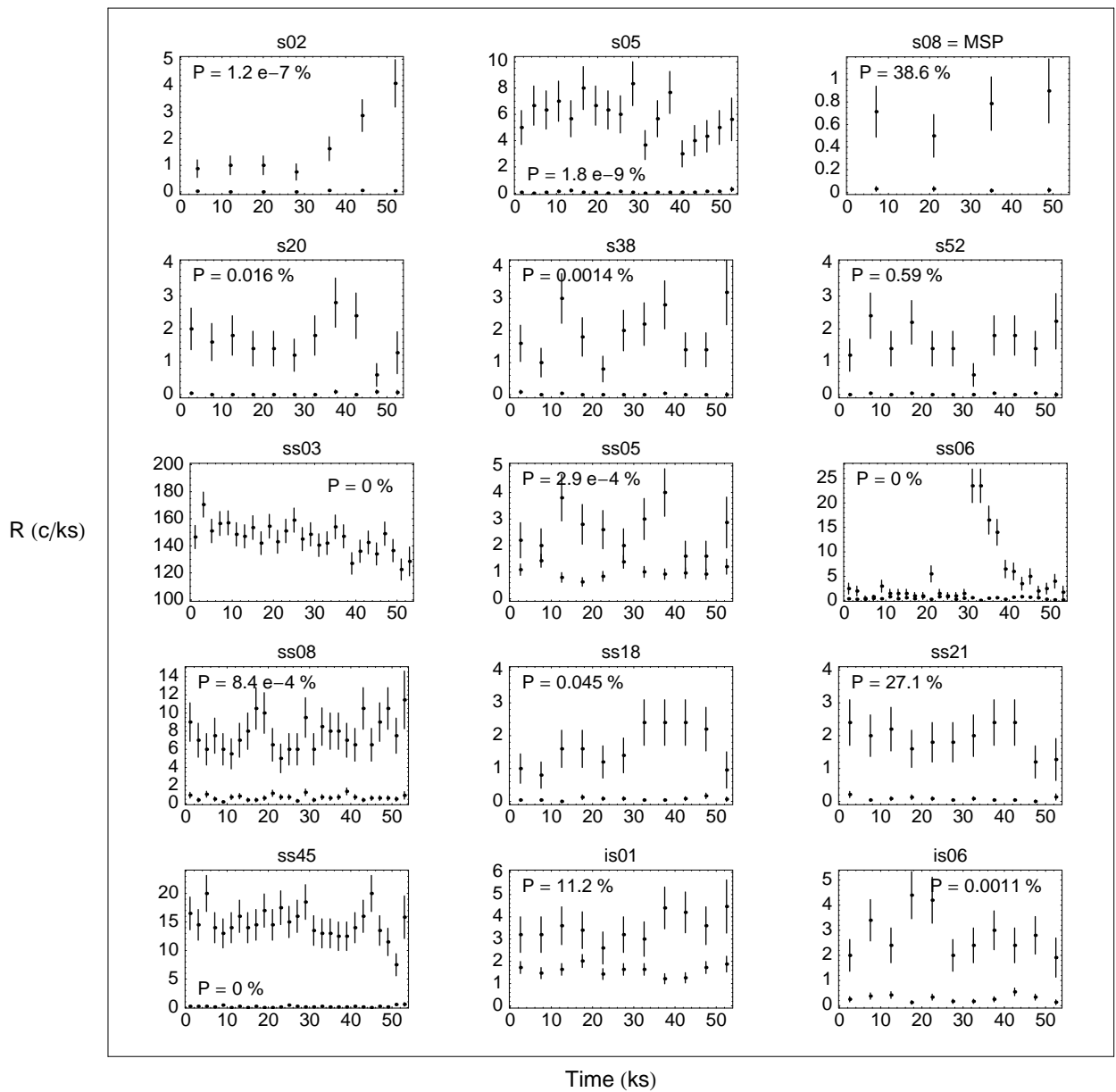}
\caption{\label{f:6}
Light curves (0.3--8.0 keV rate in c/ks vs. time in ks) for the 14 \Chandra\ detected X-ray sources in Tables~\ref{t:1}--\ref{t:2} with 0.3--8.0 keV counts $C_x > 80$, plus the X-ray source, {\it s08}, coincident with the MSP in M71.  Both the source and the local background light curves are shown.
}
\label{f:fig6}
\end{figure}


\begin{figure}[htbp]
\epsscale{1.0}
\plotone{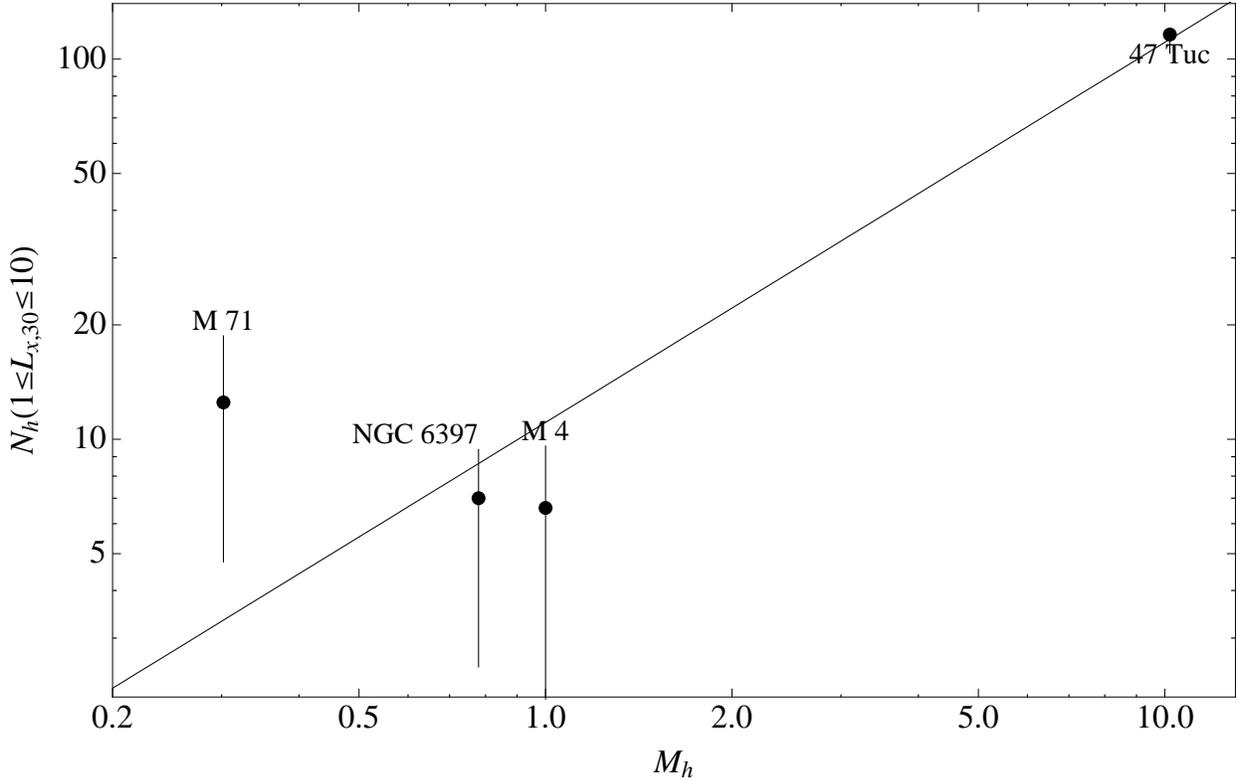}
\caption{\label{f:7}
For four clusters, number of background-subtracted sources inside $r_h$ with 0.5---2.5 keV X-ray luminosities, $L_{x,30}$, in the range 1---10, vs. scaled values for the mass, $M_h$, inside $r_h$.
Following Kong \et\ (2006), scaled values for $M_h$ were calculated from $M_h = 10^{-0.4 (M_V - M_{V,M4})}$, where $M_V$ is the cluster absolute visual magnitude from Harris (1996, updated 2003).
The value $M_{V,M4}$ for the cluster M4 is -7.2.
The result of a weighted linear fit to the data ($N_h = a M_h$), including M71, is also plotted, with slope $a = 11.1$.
}
\label{f:fig7}
\end{figure}

\clearpage

\end{document}